\documentclass[aps,pra,twocolumn,longbibliography,superscriptaddress,nofootinbib]{revtex4-2}

\usepackage[T1]{fontenc}
\usepackage{lmodern}
\usepackage{amsmath,amssymb,amsfonts,amsthm,mathtools,float,graphicx,version}
\usepackage[varg,amsthm]{newtx}
\usepackage[cal=dutchcal]{mathalpha}
\excludeversion{redact}
\usepackage{bm}
\usepackage{microtype}

\usepackage[shortlabels]{enumitem}
\setlist[enumerate,1]{1.,itemsep=2pt,parsep=0pt,topsep=3pt,left=0pt}
\setlist[itemize,1]{left=0pt}

\usepackage[colorlinks,linkcolor=blue,citecolor=blue,urlcolor=blue]{hyperref}

\newcommand{\al}{\alpha}
\newcommand{\be}{\beta}
\newcommand{\ga}{\gamma}
\newcommand{\de}{\delta}
\newcommand{\ka}{\kappa}
\newcommand{\si}{\sigma}
\newcommand{\om}{\omega}
\newcommand{\ze}{\zeta}

\renewcommand{\th}{\theta}

\newcommand{\vep}{\varepsilon}
\newcommand{\Ga}{\Gamma}
\newcommand{\Si}{\Sigma}
\newcommand{\bpsi}{\bm{\psi}}
\newcommand{\bphi}{\bm{\phi}}
\newcommand{\bPsi}{\bm{\varPsi}}

\newcommand{\bE}{\bm E}
\newcommand{\bcE}{\bm{\mathcal E}}
\newcommand{\bcB}{\bm{\mathcal B}}

\newcommand{\bF}{\bm F}
\newcommand{\bM}{\bm M}
\newcommand{\bQ}{\bm Q}
\newcommand{\bZ}{\bm Z}
\newcommand{\bS}{\bm S}
\newcommand{\bu}{\bm u}
\newcommand{\EE}{\mathbb E}
\newcommand{\RR}{\mathbb{R}}
\newcommand{\CC}{\mathbb{C}}
\newcommand{\bk}{\bm{k}}
\newcommand{\bh}{\bm{h}}
\newcommand{\bx}{\bm{x}}
\newcommand{\bn}{\bm{n}}
\newcommand{\bP}{\bm{P}}
\newcommand{\bp}{\bm{p}}
\newcommand{\bq}{\bm{q}}
\newcommand{\cF}{\bm{\mathcal F}}
\newcommand{\dd}{\mathrm{d}}
\newcommand{\pd}{\partial}
\newcommand{\pdf}[2]{\frac{\partial #1}{\partial #2}}
\newcommand{\T}{\mathsf T}
\newcommand{\cH}{\mathcal H}

\newcommand{\sgn}{\operatorname{sgn}}

\newcommand{\ket}[1]{|#1\rangle}

\newcommand{\bra}[1]{\langle #1|}

\theoremstyle{remark}
\newtheorem{remark}{Remark}

\begin{document}

\title{Photon position eigenstates in configuration space}

\author{Artemio González-López}\email{artemio@ucm.es}
\author{Luis Martínez Alonso}\email{luism@ucm.es}

\affiliation{Departamento de Física Teórica, Facultad de Ciencias Físicas, Universidad
  Complutense, 28040 Madrid, Spain}
\date{\today}
\begin{abstract}
  The expressions of the eigenfunctions of the Hawton photon position operator in configuration
  space are derived for several classes of wave function, including the Riemann--Silberstein and
  Landau--Peierls cases. Although these eigenfunctions have a simple form in momentum space, the
  explicit characterization of their representations in configuration space is rather more
  involved. We provide closed expressions of these eigenfunctions in terms of linear combinations
  of the complete elliptic integrals $K(\ka)$ and $E(\ka)$ with modulus $\ka$ depending on
  trigonometric functions of the polar angle. We show that they diverge not only at the value
  $\bq$ of the position eigenvalue, but also on a plane containing $\bq$ and that they decay as
  inverse powers of the distance from $\bq$.
\end{abstract}

\maketitle

\section{Introduction}

The development of photon wave mechanics, i.e., the first quantized theory of the photon, the
existence of a suitable photon wave function, and the understanding of the photon localization
problem are long-standing and controversial subjects of research deeply related to the existence
of a suitable photon position operator (see, e.g.,
\cite{NW49,JO78,Ha99,HB01,Ke05,BB98,BB12,De15,DPTT23}, and in particular the discussions in
\cite{Ke05}, \cite{De15}, and~\cite{DPTT23}). The main difficulty arises due to the assumption
that the coordinate wave function should be the projection $\langle \bq|\psi\rangle$ of the
quantum vector state $|\psi\rangle$ on the eigenfunctions $|\bq\rangle$ of the position operator.
Indeed, since the classical work by Newton and Wigner~\cite{NW49}, it has been frequently stated
that an appropriate photon position operator does not exist \cite {CDG89,Bo54,Po64}, and,
consequently, it is impossible to define a wave function for the photon in the coordinate
representation. For instance, D. Bohm claims in \cite{Bo54}: ``One cannot properly speak of a wave
function of the photon localized in ordinary space.'' Thus, the use of the photon wave function is
usually limited to the momentum representation, since the photon momentum operator is well
defined, and therefore the notion of photon wave function in the momentum representation is well
founded. In spite of these claims, a position operator with commuting components has been
constructed by Hawton \cite{Ha99} (see also \cite{De15,DPTT23,DPTT21}). On the other hand, the
concept of photon wave function in configuration (i.e., $\bx$) space (``Konfigurationsraum'') was
already introduced by Landau and Peierls in 1930 \cite{LP30} without any reference to a position
operator, via an inverse Fourier transform. This approach for introducing the photon wave function
in the coordinate representation, proposed (among others) by I. Bialynicki-Birula in \cite{BB05},
is the one that we shall adopt in the present work.

In this paper we investigate the configuration space versions of the eigenfunctions of the Hawton
position operator. There is a family of Hawton position operators associated to each orthonormal
basis $\{\bm{E}_1(\bk),\bm{E}_2 (\bk), \bm E_3(\bk)=\bk/|\bk|\}$ in 3D Euclidean momentum space
(see, e.g., \cite{DPTT23}).
This frame dependence is a known feature of the construction and is closely related to the
geometrical, or gauge-like, freedom discussed in previous work on localized photon
states~\cite{HB01,HB05,DPTT21}. It is also one of the reasons why the physical interpretation of
photon position operators remains delicate. In the present work we do not attempt to settle this
interpretational issue. Rather, we study the configuration-space eigenfunctions associated with
this class of operators, with special emphasis on the standard spherical frame. Thus our results
should be understood as an explicit analysis of the localization bases associated with definite
representatives of this family of photon position operators.

The Hawton position operators $\bQ$ are constructed to ensure that the states of the form
\begin{equation}\label{psij0}
  \bpsi^{(j)}_{\bq}(\bk)\propto k^{\be/2}\bm{E}_j(\bk)\,e^{-i\bk\cdot \bq},\quad j=1,2
\end{equation}
(in momentum space) are common eigenstates of the three components $Q_l$ of $\bQ$ with eigenvalue
$q_l$. The parameter $\be$ specifies the type of photon wave function used, with $\be=1, \be=0 $
and $ \be=-1$ corresponding to the Riemann--Silberstein (RS), Landau--Peierls (LP), and vector
potential wave functions, respectively~\cite{Ha07}. The $\bk$-dependent functions~\eqref{psij0}
are familiar in the semiclassical treatment of electromagnetic radiation (see, e.g.,
\cite{Sa67})). Although
the available explicit expressions of the functions~\eqref{psij0} in configuration space are
rather involved~\cite{BM17,DPTT21}, a detailed and systematic analysis of these expressions and
their properties is required in the formalism of photon wave mechanics. This formalism is closely
related to the program discussed in ref.~\cite{BB94}, where the connection between classical
Maxwell electrodynamics and quantum field theory was described, in a specific context, as a
``missing link''.

In section~\ref{sec.QMp} we discuss the position operators associated with the orthonormal basis
of the spherical coordinate system in three-dimensional space and the Hilbert spaces with weight
$|\bk|^{-\be}$ in the momentum (or wave vector $\bk$) representation. More precisely, we consider
the family of candidate wave functions given by the product of the weight $|\bk|^{-\be}$ times the
RS field~\cite{We01,Si07}. The two main wave functions in this family are the RS field and the LP
wave function. Moreover, due to the presence of the dynamical superselection rule associated with
the photon helicity, we focus on the two sectors of the Hilbert space determined by the two
helicity values.

We present in section~\ref{sec.ppo} and appendices~\ref{app.HPO}-\ref{app.norm} a self-contained
discussion of the position operators associated with general frames in 3D Euclidean momentum
space, with special emphasis in the spherical coordinate system (i.e., the standard frame). We
then turn in section~\ref{sec.poseigf} to the analysis of the eigenfunctions of these operators.
We derive explicit expressions for the eigenfunctions of the photon position operator in
configuration space, using several special functions identities provided in appendix~\ref{app.ids}
(the detailed calculations are presented in appendix~\ref{app.calc}). In particular, we provide
closed-form expressions for the LP and RS eigenfunctions of the position operator corresponding
not only to the standard frame, but also to more general frames. In addition, we derive an
explicit expression for the Hertz superpotential of the RS field at the initial time, and discuss
its time evolution (the details of the calculation appear in appendix~\ref{app.Hertz}). For the
standard frame, the eigenfunctions with well-defined helicity are of the form
\[
  R(r)\Big(P_{\rho}(\theta)\bm e_{\rho}+P_{\psi}(\theta)\bm e_{\psi}+P_{z}(\theta)\bm e_{z} \Big),
\]
where $r=|\bx-\bq|$ is the distance from the point $\bx$ to the eigenvalue $\bq$, $\theta$ is the
angle of $\bx-\bq$ with the $z$-axis, and $\{\bm e_{\rho}, \bm e_{\psi},\bm e_z\}$ are the unit
vectors of the cylindrical coordinate system in configuration space. The radial factor $R(r)$ is
proportional to $r^{-3}$ for LP eigenfunctions and to $r^{-7/2}$ for RS eigenfunctions. It is
worth noting in this respect that I. Bialynicki-Birula~\cite{BB98} characterized solutions of the
RS field equations exhibiting localization stronger than the eigenfunctions of the Hawton position
operator derived in this work (more concretely, exponential decay). Another interesting property
of the eigenfunctions derived in this work is that they are all real-valued; in particular, the RS
electromagnetic field is purely electric.

The singularities and asymptotic behavior of these eigenfunctions are analyzed in
section~\ref{sec.sing}. In particular, we show that they diverge not only at the position
eigenvalue $\bq$, but also on a plane containing it, thus reflecting the extended character of the
corresponding wave functions. We also briefly discuss the topological origin of this divergence,
and provide simple asymptotic expressions of the eigenfunctions near the singular plane.

\section{Quantum Mechanics of Photons}\label{sec.QMp}

\subsection{ The Riemann--Silberstein field}

The vacuum Maxwell equations in Gaussian units are
\begin{equation}
  \begin{aligned}\bm{\nabla} \times \bcE
    &=-\frac{1}{c} \frac{\partial
      \bcB }{\partial t}, &\qquad \bm{\nabla} \times \bcB &=\frac{1}{c} \frac{\partial
                                                              \bcE }{\partial t},\\
    \bm{\nabla} \cdot \bcE &=0,&\qquad \bm{\nabla} \cdot \bcB &=0.
  \end{aligned}
\end{equation}
They take a concise form, analogous to the Dirac equation, in terms of either complex-valued RS
field $ \cF_\pm\equiv\cF_\pm(\bx,t)$ defined by \cite{We01,Si07}
\begin{equation}\label{f}
  \cF_\si=\bcE +i\si \bcB ,\quad \si=\pm1,
\end{equation}
namely
\begin{equation}\label{m1m2}
  i\,\frac{\partial \cF_\si}{\partial t}=\si c \bm{\nabla} \times \cF_\si,\qquad
  \bm{\nabla} \cdot
  \cF_\si=0.
\end{equation}
As we shall establish below, the ``$\pm$'' sign is appropriate for dealing with positive energy
left handed/right-handed photons (helicity $\pm1$). Either RS vector establishes a connection
between the classical Maxwell theory and the quantum mechanics of photons \cite{BB13}. More
precisely, we consider the family of candidate photon wave functions (in the momentum
representation) defined in terms of the Fourier transform of the RS fields,
\begin{equation}
  \bF_\si(\bk,t)
  = \frac{1}{(2\pi)^{\frac32}} \int_{\mathbb{R}^3} \dd^3x\, \cF_\si(\bx,t)\, e^{-i \bk \cdot \bx}\,,
\end{equation}
by
\begin{equation}\label{psikal}
  \bpsi_\si(\bk,t)=(\hbar c k)^\al\bF_\si(\bk,t), \quad k=|\bk|,
\end{equation}
where $\al$ is a real parameter. The most popular choices of this parameter are $\al=0$, yielding
the RS wave function $\cF_\si$, $\al=-1/2$, which yields the Landau--Peierls~\cite{LP30} wave
function (in the momentum representation)
\begin{equation}\label{lpf}
  \bpsi_\si(\bk,t)=(\hbar ck)^{-\frac12} \bF_\si(\bk
  ,t),
\end{equation}
and $\al=-1$, corresponding to the vector potential wave function. Note that, since in the
Gaussian system of units $\cF_\si$ has dimensions of $E^{1/2}L^{3/2}$ (where
$E\equiv\text{energy}$), the dimension of $\bpsi_\si(\bk,t)$ in eq.~\eqref{psikal} is
$E^{\al+\frac12}L^{3/2}$. Thus the squared moduli $|\bpsi_\si(\bk,t)|^2$ and $|\bF_\si(\bk,t)|^2$
of the LP and RS wave functions have respective dimensions of \emph{probability density} and
\emph{energy density} in momentum (or, more precisely, wave vector) space. In fact, in the latter
case (i.e., for $\al=0$) we have
\[
  \int_{\RR^3}\dd^3k\,|\bF_\si(\bk,t)|^2=
  \int_{\RR^3}\dd^3x\,\big(|\bcE(\bx,t)|^2+|\bcB(\bx,t)|^2\big),
\]
which is the total energy of the classical EM field. In configuration space, the RS wave function
$\cF_\si(\bx,t)$ and the general wave function $\bPsi_\si(\bx,t)$ with $\al\ne0$ are non-locally
related (for $\al$ not an even non-negative integer) by the identity~\cite{BB96,BB05}
  \begin{align*}
    (\hbar c)^{-\al}\bPsi_\si(\bx,t)
    &= (-\Delta)^{\al/2}\cF_\si(\bx,t)\\
    &=
      -\frac{\Ga(\al+2)}{2\pi^2}\sin\!\left(\frac{\pi\al}2\right)
      \int_{\mathbb{R}^3} \dd^3y\,\frac{\cF_\si(\bm{y},t)}{|\bx-\bm{y}|^{\al+3}}\,.
  \end{align*}

  In terms of the Fourier-transformed RS function~$\bF_\si(\bk,t)$, the Maxwell equations
  \eqref{m1m2} take the form
\begin{equation*}
  \frac{\partial\bF_\si}{\partial t}=\si c\, \bk \times \bF_\si,\qquad
  \bk \cdot \bF_\si=0,
\end{equation*}
whence it follows that the general photon wave function~\eqref{psikal} satisfies the analogous
equations
\begin{equation}\label{l1lp}
  \frac{\partial \bpsi_\si}{\partial t}=\si c\, \bk \times \bpsi_\si,\qquad
  \bk \cdot \bpsi_\si=0
\end{equation}
for arbitrary values of the parameter $\al$.

\subsection{Photon wave mechanics}

For either choice of $\si$, we introduce the photon Hilbert space $\cH_\be$
as the linear space of functions $\bpsi:\mathbb{R}^3\mapsto \mathbb{C}^3$ such that
\begin{equation}\label{esp}
  \int_{\mathbb{R}^3} \frac{\dd^3k}{k^\be}\,\sum_{i=1}^3|\psi_i(\bk)|^2<\infty,
\end{equation}
verifying the transversality condition
\begin{equation}
  \bk \cdot \bpsi(\bk)=0,
\end{equation}
where $\be\in\RR$ is a real parameter to be determined (cf.~eq.~\eqref{beal} below). The scalar
product in $\cH_\be$ is defined by
\begin{subequations}\label{scalprod}
  \begin{equation}\label{scalprod1}
    (\bpsi,\bphi):=\int_{\RR^3}\frac{\dd^3k}{(\hbar ck)^\be}\,
    \bpsi^*(\bk)\cdot\bphi(\bk),
  \end{equation}
  where
  \begin{equation}
    \label{scalprod2}
    \bpsi^*(\bk)\cdot\bm{\phi}(\bk):=\sum_{i=1}^3
    \psi_{i}^*(\bk)\phi_{i}(\bk)
  \end{equation}
\end{subequations}
Some basic Hermitian operators on $\mathcal{H}_\be$ are:
\begin{itemize}
\item The momentum operator $\bP$, with components
  \begin{equation}\label{p}
    P_j \bpsi(\bk)=\hbar  k_j\bpsi(\bk),\quad j=1,2,3.
  \end{equation}

\item The Hamiltonian operator $H$ defined by
  \begin{equation}\label{ho}
    H\bpsi(\bk)=\hbar\om(k)\psi(\bk),\quad \om(k):=ck.
  \end{equation}
  
\item The helicity operator
  \begin{equation}\label{hel}
    \Si \bpsi(\bk)=\frac{\bm{P}\cdot \bm{S}}{P}\bpsi(\bk)
    =i \,\frac{\bk}{k} \times \bpsi(\bk),
  \end{equation}
  where $\bm{S}=(S_i)_{i=1}^3$ is the spin-one vector operator whose components are the generators
  of the fundamental representation of the $\mathrm{so}(3)$ Lie algebra:
  \begin{equation*}
    S_1=\begin{pmatrix}
      0 & 0 & 0\\
      0 & 0 & -i\\\
      0 & i & 0
    \end{pmatrix},\quad 
    S_2=\begin{pmatrix}
      0& 0 & i\\
      0 & 0 & 0\\\
      -i & 0 & 0
    \end{pmatrix},\quad
    S_3=\begin{pmatrix}
      0 & -i & 0\\
      i & 0 & 0\\\
      0 & 0 & 0
    \end{pmatrix}.
  \end{equation*}
  In other words, $(S_j)_{kl}=-i\,\vep_{jkl}$, where $\vep_{jkl}$ is Levi-Civita's completely
  antisymmetric tensor. Furthermore, since $\Si^2=I$ the eigenvalues of $\Si$ are $\pm1$, and
  their corresponding eigenstates are of the form $(1\pm\Si)\bpsi$ for arbitrary
  $\bpsi\in\mathcal H_\be$.
\end{itemize}

With the definition~\eqref{hel} of the helicity, the evolution equation \eqref{l1lp} for the
general photon wave function~\eqref{psikal} can be written as
\begin{equation}\label{Schreq}
  i\hbar \frac{\partial \bpsi_\si}{\partial t}=\si\hbar\om(k)\Si\bpsi_\si.
\end{equation}
This equation takes the form of the Schrödinger equation for the Hamiltonian~\eqref{ho} provided
that
\begin{equation}\label{helcond}
  \Si\bpsi_\si=\si\bpsi_\si,
\end{equation}
i..e, that $\bpsi_\si$ is a helicity eigenstate with eigenvalue $\si$. In other words, photons
with energy $\hbar\om(k)$ and helicity $\si$ are described by wave functions $\bpsi_\si$
determined by the RS vector $\cF_\si=\bcE+i\si\bcB$ satisfying eq.~\eqref{helcond}. Thus, for
$\si=1$ the wave function $\bpsi_\si$ and equations~\eqref{Schreq}-\eqref{helcond} describe
photons with helicity $+1$, while for $\si=-1$ they describe photons with helicity $-1$, with
positive energy $\hbar\om(k)$ regardless of the helicity.

In fact, since in the case of massless particles the helicity operator $\Si$ commutes with the
generators of the Poincaré algebra, the two helicity eigenspaces are invariant under
\emph{continuous} (i.e., proper orthochronous) Poincaré transformations. If we restrict ourselves,
as we shall do in this work, to these transformations, helicity acts as an effective
superselection rule forbidding the linear superposition of states with different helicities. This
is fully consistent with our construction of the photon's wave function, since the sign of the
helicity fixes how the wave function is related to the electromagnetic field vectors through
equations~\eqref{f}-\eqref{psikal}. It should be noted, however, that space reflections reverse
the sign of the helicity. Hence helicity is \emph{not} a fundamental superselection rule, but
rather a \emph{dynamical} one. In other words, superpositions of states with different helicities
are not detectable in practice, since fundamental physical observables like energy and momentum
commute with the helicity operator\footnote{More precisely, a linear combination
  $a_+\bpsi_++a_-\bpsi_-$, with $a_\pm\in\CC$ and $|a_+|^2+|a_-|^2=1$, of two photon wave
  functions $\bpsi_\pm$ with different helicities $\si=\pm1$ would be indistinguishable from the
  statistical mixture $|a_+|^2\ket{\bpsi_+}\bra{\bpsi_+}+|a_-|^2\ket{\bpsi_-}\bra{\bpsi_-}$.}

The Hilbert space $\cH_\be$ decomposes as the direct sum $\cH_\be=\cH_{\be,+}\oplus \cH_{\be,-}$
of the subspaces $\cH_{\be,\pm}$ of helicity eigenstates with eigenvalues $\si=\pm1$, and
admissible pure states cannot be superpositions of states with different helicities $\pm1$. In
fact, if $\bpsi(\bk)$ is an arbitrary element of $\cH_\be$ the function
\begin{equation}\label{eq2}
  \bpsi_{\si}(\bk)=\bpsi(\bk)+\si   \Si  \bpsi(\bk)
\end{equation}
is an helicity eigenfunction with eigenvalue $\si$. By eq.~\eqref{Schreq}, this photon state
evolves in time as
\begin{equation}\label{evol}
\bpsi_\si(\bk,t)=e^{-i\om(k)t}\bpsi_\si(\bk),
\end{equation}
with a \emph{positive} frequency $\om(k)=ck$ regardless of the helicity $\si$. Moreover, due to
the idempotent character of $\Si$ we have
\begin{equation}
 \bpsi^*_{+}(\bk,t)\cdot\bpsi_{-}(\bk,t)=\bpsi_{+}^*(\bk)\cdot\bpsi_{-}(\bk) =0.
\end{equation}
Hence there is no interference between the two photon wave functions $\bpsi_{\pm}(\bk,t)$.

The parameter $\be$---and, hence, the scalar product~\eqref{scalprod}---is determined from $\al$
by the requirement that the expectation value of the energy coincides with the total energy of the
associated classical electromagnetic field, namely
\begin{equation}\label{me}
  (\bpsi_\si,H\bpsi_\si)=\int_{\RR^3}\dd^3x\,\big(|\bcE(\bx,t)|^2+|\bcB(\bx,t)|^2\big).
\end{equation}
Indeed, by eqs.\eqref{psikal} and~\eqref{Schreq}, together with Plancherel's theorem, we have
\begin{align*}
  (\bpsi_\si,H\bpsi_\si)
  &=\int_{\RR^3}\frac{\dd^3k}{(\hbar ck)^\be}\,(\hbar ck)^{1+2\al}
                      |\bF_\si(\bk,t)|^2\\
  &=\int_{\RR^3}\dd^3k\,|\bF_\si(\bk,t)|^2=\int_{\RR^3}\dd^3x\,|\cF_\si(\bx,t)|^2
\end{align*}
provided that
\begin{equation}\label{beal}
  \be=1+2\al.
\end{equation}
Each value of $\al$ thus defines a different quantum model---in particular, a different Hilbert
space $\cH_{1+2\al}$ via the scalar product~\eqref{scalprod}---through the relation~\eqref{psikal}
connecting the photon wave function with the classical electromagnetic field.
For $\al=0$ (RS wave function) we have $\be=1$, which yields the Lorentz-invariant
Bialynicki-Birula scalar product, while the choice $\al=-1/2$ (LP wave function) leads to the
standard (but not covariant) scalar product
\begin{equation*}
  (\bpsi,\bphi)=\int_{\RR^3}\dd^3k\,\bpsi^*(\bk)\cdot\bphi(\bk).
\end{equation*}
These are also the scalar products considered in ref.~\cite{Ha07} (cf.~eq.~(6) and the preceding
unnumbered equation), where photon wave functions in momentum space are derived by projection onto
a basis of position eigenvectors. Another common choice of the parameter $\al$ is $\al=-1$ (vector
potential wave function), or equivalently $\be=-1$, with corresponding scalar product
\[
  (\bpsi,\bphi)=\hbar c\int_{\RR^3}\dd^3k\, k\bpsi^*(\bk)\cdot\bphi(\bk).
\]

\subsection{The Hertz superpotential of the RS field}
As shown in~\cite{BB98} and~\cite{BB13}, in order to characterize explicit solutions of the
Maxwell equations it is useful to represent the RS field in the form
\begin{equation}\label{he1}
  \cF_\si(\bx,t)=  \bm{\nabla} \times \Big(  (i/c)\partial_t
  \bm {Z}(\bx,t)+ \si \bm{\nabla} \times \bm {Z}(\bx,t)  \Big),
\end{equation}
where $\bZ$ is the complex (magnetic) Hertz superpotential, which can be chosen to satisfy the
wave equation
\begin{equation}\label{he 2}
 \Big(\frac{1}{c^2}\partial_{tt}-\Delta \Big) \bZ=0.
\end{equation}
The general solution of this equation can in turn be written as
\begin{equation}\label{he4}
 \bm {Z}(\bx,t)=(2\pi)^{-\frac32}\int_{\RR^3}\dd^3k\,\bh(\bm{k})e^
  {i(\bm{k\cdot \bm{x}-\omega t)}},
\end{equation}
where $\bh:\RR^3\to\CC^3$ is the Fourier transform of $\bZ(\bx,0)$ and $\om=ck$. Substituting
this equation into eq.~\eqref{he1} we then obtain the following formula for the general solution
of the evolution equation for the RS wave function:
\begin{equation}\label{he3}
  \cF_\si(\bx,t)=\int_{\RR^3}\frac{\dd^3k}{(2\pi)^{\frac32}}
  \,\bm{k}\times\Big(ik\bh(\bm{k})
    -\sigma\bm{k}\times \bh(\bm{k})\Big)e^{i(\bm{k}\cdot \bm{x}-\omega t)}.
  \end{equation}
  Note that in order for $\cF_\si$ to be square integrable we must require that
  $k\bk\times\bh(\bk)\in\cH_1$.
 
\begin{remark}
  So far, we have simultaneously dealt with photons of either helicity~$\si=\pm1$ on the same
  footing. To simplify the notation, from now on we shall drop the subindex $\si$
    from the wave function.\qed
\end{remark}
 
\section{Photon position operator}\label{sec.ppo}

A position operator $\bm{Q}=(Q_i)_{i=1}^3$ for the photon must satisfy the following fundamental
properties~\cite{DPTT23}:
\begin{enumerate}
\item The components $Q_i$ are Hermitian operators on $\cH_\be$. In particular,
  $ \bk \cdot Q_i\bpsi(\bk)=0$ for all $\bpsi\in\mathcal H$ and $i=1,2,3$.
\item The components $Q_i$ obey the canonical commutation relations
  \begin{equation}\label{commrel} [Q_i,Q_j]=0 , \quad
    [Q_i,P_j]=\hbar \delta_{ij},\qquad \forall i,j=1,2,3.
  \end{equation}

\item Each $Q_i$ commutes with the helicity operator $\Si$.
\end{enumerate}

As shown in ref.~\cite{Ha99} (see ref.~\cite{DPTT21} for an alternative derivation), a position
operator can be constructed from any orthonormal right-handed frame $\{\bm{E}_i(\bk)\}_{i=1}^3$
on $\mathbb{R}^3 \setminus\{(0,0,k_3)\mid k_3\geq 0\}$ such that
\begin{equation}\label{ob}
  \bm{E}_i(\bk)\cdot \bm{E}_j(\bk)=\delta_{ij},
  \quad \bm{E}_1(\bk)\times \bm{E}_2(\bk)=\bm{E}_3(\bk)=\frac{\bk}{k}.
\end{equation}
More precisely, if we define the real orthonormal matrix $\mathbb{E}$ with matrix elements given
by $\mathbb{E}_{ij}=(\bm{E}_j)_i$ it is straightforward to check that,
if we adopt the scalar
product~\eqref{scalprod},  a position operator verifying the above properties  turns out to be
\begin{equation}\label{RSQj}
  Q_j=i\,k^{\be/2}\,\mathbb E\,\frac{\partial}{\partial k_j}\,\mathbb E^{-1}k^{-\be/2}, \quad j=1,2,3.
\end{equation}

In what follows, we shall assume unless otherwise stated that the frame $(\bm{E}_i(\bk))_{i=1}^3$
on $\mathbb{R}^3 \setminus\{(0,0,k_3)\mid k_3\geq 0\}$ is provided by the orthonormal basis
associated to the set of spherical coordinates $(k,\theta,\phi)$ on $\mathbb{R}^3$:
\begin{subequations}\label{basis}
\begin{align}
  \bm{E}_1(\bk)&=\cos\theta\,\cos\phi \,\bm{e}_1+ \cos\theta\,\sin\phi\, \bm{e}_2
                    -\sin\theta \,\bm{e}_3\notag\\
                    &=\frac{1}{kk_{\perp}}\bk\times (\bk\times\bm{e}_3),\label{bE1}\\
  \bm{E}_2(\bk)&=-\sin\phi \,\bm{e}_1+ \cos\phi \,\bm{e}_2
                    =-\frac{1}{k_{\perp}}\bk\times\bm{e}_3,
  \label{bE2}\\
  \bm{E}_3(\bk)&=\sin \theta\,\cos\phi \,\bm{e}_1
                 + \sin\theta\,\sin\phi \,\bm{e}_2+\cos \theta\, \bm{e}_3=\frac{\bk}{k},
                 \label{bE3}
\end{align}
\end{subequations}
where $(\bm{e}_i)_{i=1}^3$ is the canonical basis of $\mathbb{R}^3$ and
\[
  k_{\perp}:=(k_1^2+k_2^2)^{\frac12}.
\]
This is also the frame used by Hawton~\cite{Ha99,Ha07}, Debierre~\cite{De15}, and Babaei and
Mostafazadeh~\cite{BM17}, among others. Of course, any other such frame $(\bm{E}'_i(\bk))_{i=1}^3$
is obtained from \eqref{basis} by a transformation of the form
\begin{equation}\label{Eiprime} 
  \bm{E}'_1=a\, \bm{E}_1 -b \, \bm{E}_2,\quad
  \bm{E}'_2=b\,\bm{E}_1 +a \, \bm{E}_2 ,\quad
  \bm{E}'_3=\bm{E}_3,
\end{equation}
where $ a=a(\bk)$ and $b=b(\bk)$ are real-valued functions such that $a^2+b^2=1$. For instance,
the frame used in ref.~\cite{DPTT21} correspond to the choice $(a,b)=(k_1,k_2)/k_\perp$.

In particular, for $\alpha=-1/2$---i.e., for the LP photon Hilbert space---the position operator
\eqref{RSQj} associated with the basis \eqref{basis} reduces to
\begin{equation}\label{po}
  Q_j=i \,\mathbb{E}\,\frac{\partial}{\partial k_j}\,\mathbb{E}^{-1}, \quad j=1,2,3.
\end{equation}
The Landau--Peierls position operator~\eqref{po} can be expressed as a $3\times3$ matrix with
elements given by the scalar operators
\begin{equation}\label{H1}
  \bQ_{nm}
  =i \delta_{nm}\bm{\nabla}_{\bk}-i \sum_{l=1}^3 (\bm{\nabla}_{\bk} \mathbb{E}_{nl}) \mathbb{E}_{ml}.
\end{equation}
If we write the last term using the spin-one operator $\bm{S}$, we obtain the formula for the
position operator in eq.~(17) of ref.~\cite{Ha99}:
\begin{subequations}\label{H2all}
  \begin{equation}\label{H2}
    \bm{Q}=i \bm{\nabla}_{\bk}+\frac{1}{k^2}\,\bk\times
    \bm{S}-\frac{1}{k} \cot\theta\,\bm{E}_2\,(\mathbb{E}S_3\mathbb{E}^{-1})
  \end{equation}
  (see appendix~\ref{app.HPO} for a simple proof). In fact, the last term of the previous formula
  turns out to be closely related to the Berry phase for photons \cite{Ha99,HB01}. Note also that
  an equivalent expression for the position operator~\eqref{H2} is
  \begin{equation}\label{H2alt}
    \bm{Q}=i \bm{\nabla}_{\bk}+\frac{1}{k^2}\,\bk\times
    \bm{S}-\frac{1}{k} \cot\theta\,\bm{E}_2\,\Si
  \end{equation}
\end{subequations}
(cf.~appendix~\ref{app.HPO}).

For arbitrary values of the parameter $\be$ (or, equivalently, $\al$), combining eqs.~\eqref{RSQj}
and~\eqref{H2all} we obtain the more general formula
\begin{subequations}\label{Qbetaall}
  \begin{equation}\label{QbetaS3}
    \bm{Q}=i \bm{\nabla}_{\bk}-\frac{\be\bk}{2k^2}+\frac{1}{k^2}\,\bk\times
    \bm{S}-\frac{1}{k} \cot\theta\,\bm{E}_2\,(\mathbb{E}S_3\mathbb{E}^{-1}),
  \end{equation}
  or equivalently
  \begin{equation}\label{QbetaSi}
    \bm{Q}=i \bm{\nabla}_{\bk}-\frac{\be\bk}{2k^2}+\frac{1}{k^2}\,\bk\times
    \bm{S}-\frac{1}{k} \cot\theta\,\bm{E}_2\,\Si.
  \end{equation}
\end{subequations}
In particular, setting $\be=1$ yields the position operator for the RS wave function.

\section{Photon position eigenfunctions}\label{sec.poseigf}

\subsection{Photon position eigenfunctions in the momentum representation}

In the momentum representation, the photon position eigenfunction $\bpsi_{\bq}(\bk)$ with
position $\bq=(q_{1},q_{2} ,q_{3})\in\mathbb{R}^3$ satisfies the system of eigenvalue
equations
\begin{equation}\label{pop}
  Q_j\bpsi_{\bq}(\bk)= q_{j}\, \bpsi_{\bq}(\bk), \qquad j=1,2,3,
\end{equation}
By eq.~\eqref{RSQj}, these equations are equivalent to the system of differential equations
\[
  i\,\frac{\partial}{\partial k_j}\,(k^{-\be/2}\mathbb{E}^{-1}
  \bpsi_{\bq})=q_{j}\,(k^{-\be/2}\mathbb{E}^{-1} \bpsi_{\bq}), \quad j=1,2,3.
\]
The general solution of this system is of the form
\begin{equation}\label{gs}
  \bpsi_{\bq}(\bk)=(\hbar ck)^{\be/2}(\mathbb{E}\,\bm{c})\, e^{-i\bk\cdot \bq},
\end{equation}
where $\bm{c}=(c_i)_{i=1}^3$ is an arbitrary constant vector in $\mathbb{C}^3$. Note that, as
explained below, we have explicitly included the factor $(\hbar c)^{\be/2}$ so that the
integration constants $c_i$ are dimensionless. Furthermore, as
$\bk\cdot \bpsi_{\bq}=0$ we must take $c_3=0$. We thus have
\begin{equation}\label{RSpsi}
  \bpsi_{\bq}(\bk)=c_1\bpsi^{(1)}_{\bq}(\bk)+c_2\bpsi^{(2)}_{\bq}(\bk),
\end{equation}
where
\begin{equation}\label{psij}
  \bpsi^{(j)}_{\bq}=(\hbar ck)^{\be/2}\bm{E}_j(\bk)\,e^{-i\bk\cdot \bq},\quad j=1,2.
\end{equation}
These states satisfy the delta function normalization
\begin{equation}\label{psinorm}
  (\bm{\psi^{(j)}}_{\bq},\bpsi^{(j')}_{\bm{q}'})=(2\pi)^3\de_{jj'}\delta(\bq-\bm{q}')
\end{equation}
with respect to the general scalar product defined in eq.~\eqref{scalprod}. Note that, by
eq.~\eqref{psij}, the eigenfunctions $\bpsi_{\bq}^{(j)}$ have the dimensions of $E^{\be/2}$,
which is of course consistent with eq.~\eqref{psinorm}.

To obtain position eigenstates~$\bpsi_{\bq,\si}(\bk)$ with well-defined helicity $\si$ we note
that from eq.~\eqref{hel} and the definition~\eqref{basis} of the vectors $\bE_j(\bk)$ it follows
that
\[
  \Si\bE_1(\bk)=i\bE_2(\bk),\qquad \Si\bE_2(\bk)=-i\bE_1(\bk).
\]
Hence the vectors
\begin{equation}
  \label{heleig}
  \bu_\si(\bk)=\frac1{\sqrt2}\big(\bE_1(\bk)+i\si\bE_2(\bk)\big)
\end{equation}
are eigenvectors of the helicity operator $\Si$ with eigenvalue $\si$, satisfying the
orthogonality condition
\begin{equation}
  \label{orthu}
  \bu^*_\si(\bk)\cdot\bu_{\si'}(\bk)=\de_{\si\si'}.
\end{equation}
Thus the general solution of eq.~\eqref{pop} with helicity $\si$ is proportional to
\begin{align}
  \bpsi_{\bq,\si}(\bk)&=(\hbar ck)^{\frac\be2}e^{-i\bk\cdot\bq}\bu_\si(\bk)\notag\\
  &=\frac1{\sqrt2}\left(\bpsi_{\bq}^{(1)}(\bk)+i\si\bpsi_{\bq}^{(2)}(\bk)\right).
   \label{posheleig}
\end{align}
By eq.~\eqref{orthu}, these functions obey the orthogonality relations
\begin{equation}\label{psinorm2}
  (\bpsi_{\bq_0,\si},\bpsi_{\bq_1,\rho})=(2\pi)^3\de_{\si\rho}\delta(\bq_0-\bm{q}_1).
\end{equation}

It is also worth noting that, if
\[
\bpsi(\bk) = \bpsi_{\bq,\si}(\bk)+  \bpsi_{\bq',\si}(\bk)
\]
is linear superposition of two photon position eigenstates with the same helicity $\si\in\{\pm1\}$
and $\bq\ne\bq'$, from eqs.~\eqref{orthu}-\eqref{posheleig} it follows that
\begin{equation*}
 \bpsi^*_{\bq,\si}(\bk)\cdot   \bpsi_{\bq',\si}(\bk)
  =(\hbar ck)^{\be}e^{-i\bk\cdot(\bq-\bq')},
 \end{equation*}
 and thus $|\bpsi_{\bq,\si}(\bk)|=|\bpsi_{\bq',\si}(\bk)|=(\hbar ck)^\be$. Therefore the
 (relative) probability density for finding a value $\bk$ of the momentum is given by
\[
|\bpsi(\bk)|^2= 4(\hbar ck)^{\be} \cos^2\left(\frac{kd}{2} \cos \th \right),
\]
where $d:=|\bq-\bq'|$ and $\theta$ is the angle between $\bk$ and $\bq-\bq'$. This equation thus
accounts for the classical Young interference pattern in a double-slit--type experiment, with
point-like (infinitely narrow) slits located at $\bq$ and $\bq'$, as arising from the superposition
of single-photon position eigenfunctions with the same helicity (polarization);
cf.~ref.~\cite{Ma02}.

\subsection{Photon position eigenfunctions in configuration space}\label{sec.eigx}
  
In configuration space, the general photon position eigenfunctions \eqref{posheleig}
  with well-defined helicity $\si$ are given by the inverse Fourier transform of
  eq.~\eqref{posheleig}:
\begin{subequations}\label{bPsisi}
  \begin{align}
    \bPsi_{\bq,\si}
    (\bx)
    &= \int_{\mathbb{R}^3} \frac{\dd^3k}{(2\pi)^{\frac32}}\,\frac1{\sqrt2}
      \left(\bpsi^{(1)}_{\bq}(\bk)+i\si\bpsi^{(2)}_{\bq}(\bk)\right)e^{i \bk \cdot \bx}\notag\\
    &=\frac1{\sqrt2}\left(\bPsi_{\bq}^{(1)}(\bx)+i\si \bPsi_{\bq}^{(2)}(\bx)\right),
                       \label{cr}
  \end{align}
  where
  \begin{align}
    \bPsi_{\bq}^{(j)}(\bx)
    &=(2\pi)^{-\frac32}\int_{\mathbb{R}^3} \dd^3k\,\bpsi_{\bq}^{(j)}(\bk)e^{i \bk \cdot\bx}
      \notag\\
    &=(2\pi)^{-\frac32}\mkern-6mu
      \int_{\mathbb{R}^3} \dd^3k\,(\hbar ck)^{\be/2}\bm{E}_j(\bk)\,
      e^{i \bk \cdot (\bx - \bq)}\notag\\
    &=\bPsi_{\bm{0}}^{(j)}(\bx-\bq),\qquad j=1,2.
      \label{crj}
  \end{align}
\end{subequations}
From the real character of $\bE_{1,2}(\bk)$ and their behavior under the parity transformation
$\bk\mapsto-\bk$ it immediately follows that $\bPsi_{\bq}^{(1)}$ is real and $\bPsi_{\bq}^{(2)}$
is pure imaginary, which in turn implies that $\bPsi_{\bq,\si}$ is real for $\si=\pm1$. Note also
that the functions~$\bPsi_{\bq}^{(j)}(\bx)$, although of course do not have well-defined helicity,
are nonetheless eigenfunctions of the position operator $\bQ$ with eigenvalue $\bq$.
\begin{remark}
  The wave functions~\eqref{cr} are not mutually orthogonal unless $\be=0$ (or, more generally, an
  even natural number), due to the non-standard nature of the scalar product~\eqref{scalprod}.
  More precisely, for $\be=0$ Plancherel's theorem implies that
  \begin{subequations}\label{norm0}
    \begin{equation}\label{norm0j}
      \big(\,\bPsi_{\bq}^{(j)},\bPsi_{\bq'}^{(j')}\big)=(\bpsi_{\bq}^{(j)},\bpsi_{\bq'}^{(j')})
      =(2\pi)^3\de_{jj'}\de(\bq-\bq'),
    \end{equation}
    and hence
    \begin{equation}\label{normosi}
      \big(\,\bPsi_{\bq,\si},\bPsi_{\bq',\si'}\big)=\frac12(2\pi)^3(1+\si\si')\de(\bq-\bq').
    \end{equation}
  \end{subequations}
  On the other hand, for $\be\ne0$ we have
  \begin{subequations}\label{norm1}
    \begin{multline}\label{norm1j}
      \big(\,\bPsi_{\bq}^{(j)},\bPsi_{\bq'}^{(j')}\big)\\=-\frac{4\pi(\hbar
        c)^\be}{|\bq-\bq'|^{\be+3}}\,\sin\left(\tfrac{\pi\be}2\right)\Ga(\be+2)\,
      \de_{jj'},\quad \be\ne0,
    \end{multline}
    and therefore
    \begin{multline}\label{norm1si}
      \big(\,\bPsi_{\bq,\si},\bPsi_{\bq',\si'}\big)\\=-\frac{2\pi(\hbar
        c)^\be}{|\bq-\bq'|^{\be+3}}\,\sin\left(\tfrac{\pi\be}2\right)\Ga(\be+2)\,
      (1+\si\si'),\quad \be\ne0,
    \end{multline}
  \end{subequations}
  (see appendix~\ref{app.norm} for the details).\qed
\end{remark}

We shall next present the explicit expressions for the LP ($\be=1$) and RS ($\be=0$) position
eigenfunctions $\bPsi_{\bm 0,\si}(\bx)$ (the detailed calculations are given in
appendix~\ref{app.calc}). In both cases, we can express these eigenfunctions as
\begin{equation}\label{Psi0gen}
  \bPsi_{\bm 0,\si}(\bx)=(\hbar c)^{\frac\be2}r^{-3-\frac\be2}\left(P_\rho(\th)\bm e_\rho+\si
    P_\psi(\th)\bm e_\psi+P_z(\th)\bm e_z\right),
\end{equation}
where $r=|\bx|$, $\th$ is the angle between the vector $\bx$ and the $\bm e_3$ axis, and
$\{\bm e_\rho,\bm e_\psi,\bm e_z=\bm e_3\}$ is the standard orthonormal frame of the
$(\rho,\psi,z)$ cylindrical coordinate system in configuration space.

\begin{figure}[h]
  \centering
  \includegraphics[width=.75\linewidth]{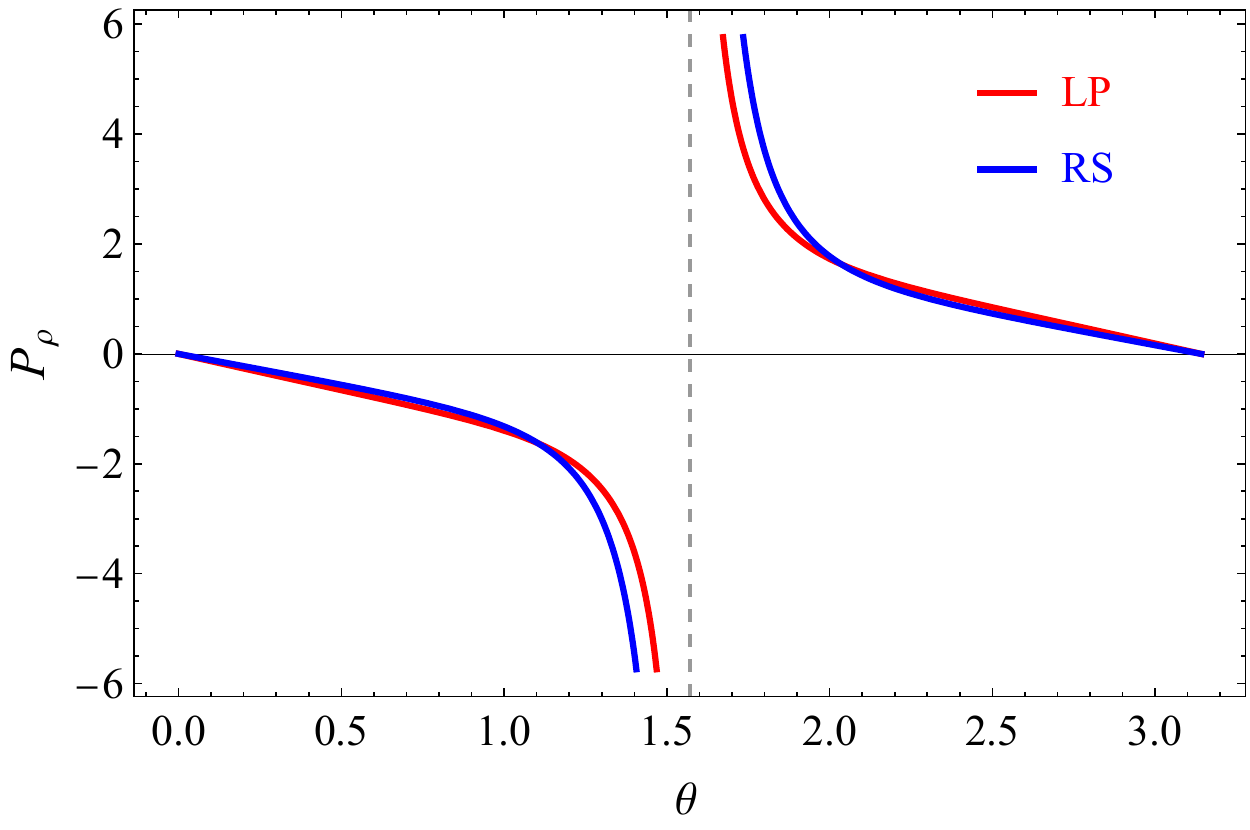}\\
    \includegraphics[width=.75\linewidth]{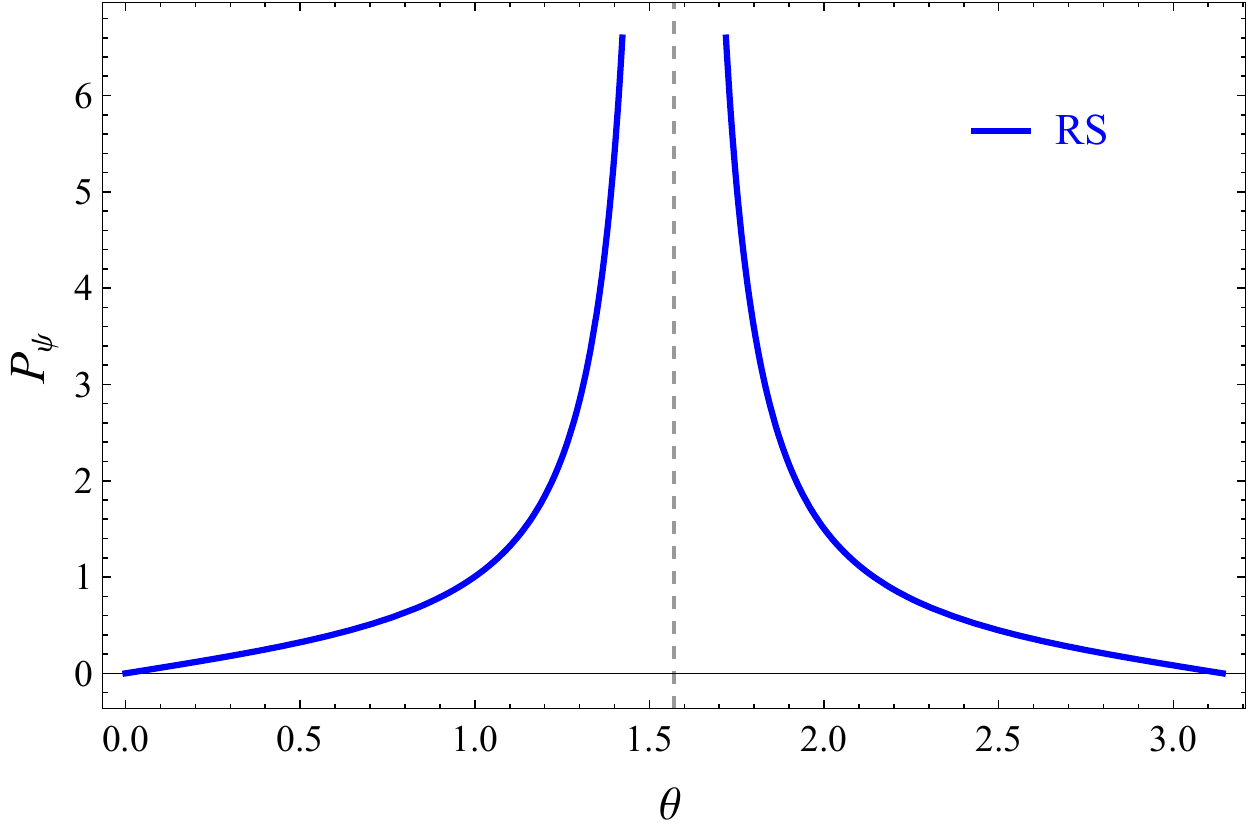}\\
  \includegraphics[width=.75\linewidth]{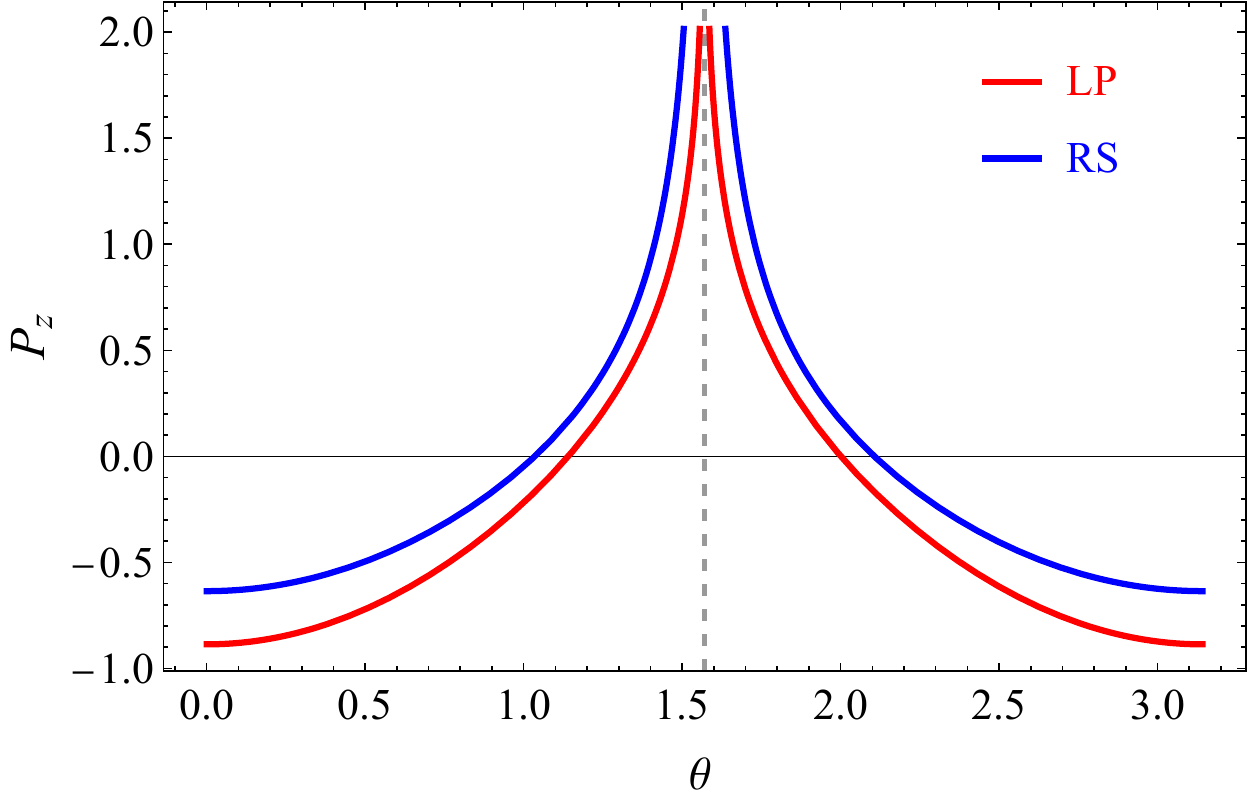}
  \caption{Plots of the functions $P_{\rho}(\th)$, $P_\psi(\th)$, and $P_z(\th)$ for the LP and RS
    photon position eigenfunctions centered at the origin (cf.~eqs.~\eqref{LPPs}-\eqref{RSPs}).
    The vertical dashed line in these plots is the singular line $\th=\pi/2$.}
  \label{fig.plots}
\end{figure}

\begin{figure}[h]
  \centering
  \includegraphics[height=6cm]{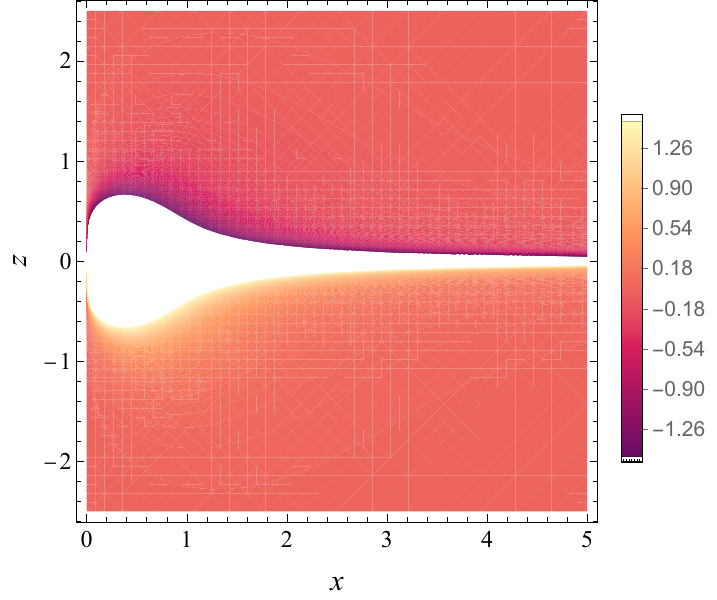}\\
    \includegraphics[height=6cm]{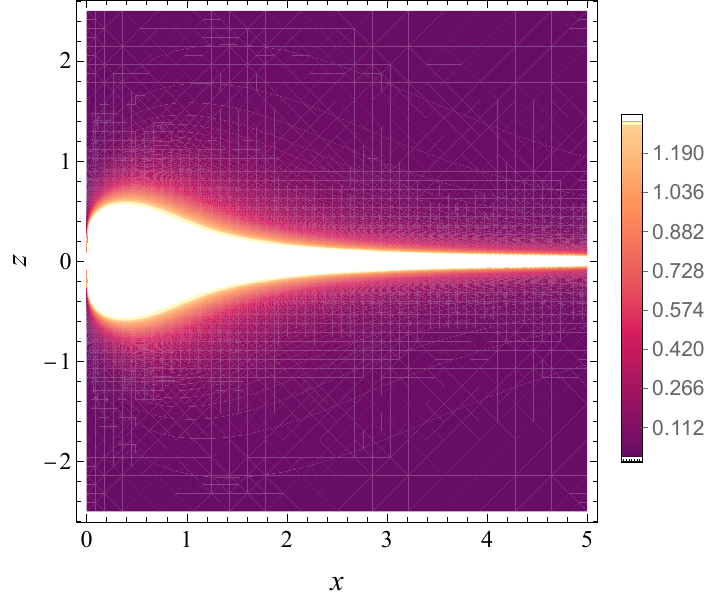}\\
  \includegraphics[height=6cm]{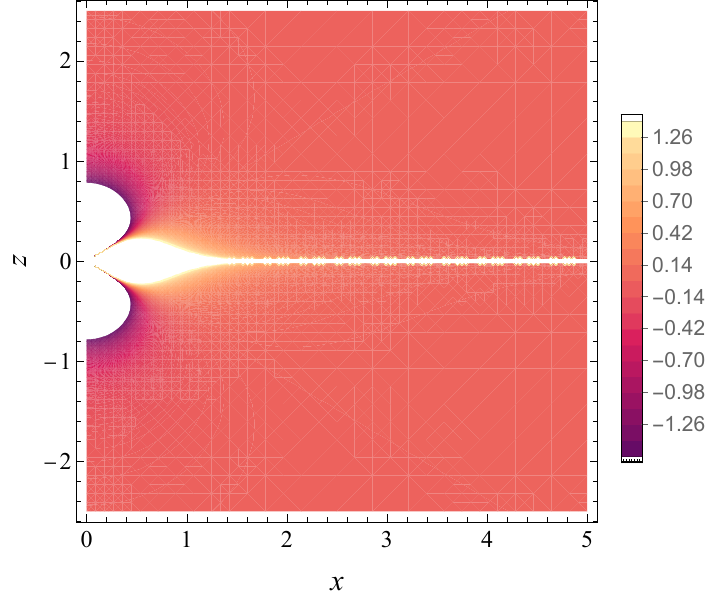}
  \caption{From top to bottom: contour plots in the $(x,z)$ plane of the $\rho$, $\psi$, and $z$
    components of the RS photon position eigenfunction with positive helicity centered at the
    origin, in units of $(\hbar c)^{1/2}$. (The white area indicates the region where the absolute
    value of these functions is greater than $1.5$.)}
  \label{fig.cplots}
\end{figure}

\begin{enumerate}
  
\item LP position eigenfunctions:
  \begin{subequations}\label{LPPs}
    \begin{align}
      \label{LPPrho}
      P_\rho&=\pi^{-\frac12}\,\big(2E(\ka_0)\cot(2\th)
              -K(\ka_0)\cot\th\big),\\
      P_\psi&=-\pi^{\frac12}\,\de\bigl(\th-\tfrac\pi2\bigr),
              \label{LPPpsi}\\
      P_z&=\pi^{-\frac12}\,\big(K(\ka_0)-2E(\ka_0)\big)
      \label{LPPz}
    \end{align}
  \end{subequations}
  where $K(\ka_0)$ and $E(\ka_0)$ are the complete elliptic integrals with modulus $\ka_0=\sin\th$
  (cf.~appendix~\ref{app.ids}).

\item RS position eigenfunctions:
  \begin{subequations}\label{RSPs}
    \begin{align}
       \label{RSPrho}
      P_\rho&= \frac{(2s)^{-\frac32}}{2K(\frac1{\sqrt2})}\,\sgn\bigl(\th-\tfrac\pi2\bigr)\,
              \frac{A_1K(\ka_1)+B_1E(\ka_1)}{\sqrt{1-s}},\\
      P_\psi&= \frac{K(\tfrac1{\sqrt2})}{\pi}\,(2s)^{-\frac32}\,
              \frac{A_2K(\ka_1)+B_2E(\ka_1)}{\sqrt{1-s}},
              \label{RSPpsi}\\
      P_z&=\frac{(2s)^{-\frac12}}{4K(\frac1{\sqrt2})}\,
           \frac{A_3K(\ka_1)+B_3E(\ka_1)}{\sqrt{1+s}},
        \label{RSPz}   
    \end{align}
    where
    \[
      s=|\cos\th|,
    \]
    $\sgn(x)=x/|x|$ is the sign function,
    \begin{equation}
      \label{ABtAtB1}
      \begin{aligned}
        A_1&=(1+s^{\frac12})(2-4s^{\frac12}+10s^2-5s^{\frac52}),\\
        B_1&=2s^{\frac12}(2-5s^2),\\
        A_2&=(3s^{\frac32}-2)(1+s^{\frac12}),\\
        B_2&=2(2-3s^2),\\
        A_3&=4+5s^{\frac12}(1+s^{\frac12}+s-s^{\frac32}),\\
        B_3&=-10s^{\frac12}(1+s),
      \end{aligned}
    \end{equation}
  \end{subequations}
  and the modulus of the elliptic integrals is given by
  \begin{equation}\label{ka1}
    \ka_1=\frac{1-\sqrt s}{\sqrt{2(1+s)}}
  \end{equation}
  The coefficients $P_i(\theta)$ are in this case linear combinations of the complete elliptic
  integrals $K$ and $E$, with modulus $\ka_1(\th)$ given by eq.~\eqref{ka1}. Note also that, for
  both the LP and RS eigenfunctions, the functions $P_i(\th)$ become singular on the horizontal
  plane $\theta=\pi/2$; see section~\ref{sec.sing} for further details. This fact is apparent from
  fig.~\ref{fig.plots}, where these functions are plotted for both the RS and LP eigenfunctions.
  The full components $r^{-3-\frac\be2}P_i(\th)$, with $i=\rho,\psi,z$, of these eigenfunctions,
  which are obviously symmetric about the $z$ axis, also become singular near the origin. To
  illustrate this behavior, we have presented in fig.~\ref{fig.cplots} contour plots of these
  components on the vertical plane $y=0$ for the positive-helicity RS eigenfunction.
  
\item Hertz potential $\bZ(\bx,t)$:

\begin{equation}
  \label{Zx0}
  \bZ(\bx,0)=-\si\bigg(\frac{\hbar c}{2}\bigg)^{\frac12}r^{-\frac32}s^{-\frac12}(1+s)^{-\frac12}
  \frac{K(\ka_1)}{K(\frac1{\sqrt2})}\,\bm e_3,
\end{equation}
with $s$ and $\ka_1$ as above. At time $t>0$, $\bZ(\bx,t)$ is of the form
\[
  \bZ(\bx,t)=-\si(\pi\hbar c)^{\frac12}\ze(\bx,t)\,\bm e_3.
\]
We have not been able to derive a closed-form expression for the scalar function $\ze(\bx,t)$ at
an arbitrary time $t$. However, its Taylor expansion about $t=0$ can be computed as outlined in
appendix~\ref{app.Hertz}, with the result
\begin{widetext}
\begin{multline}
  \label{Zseries}
   \left(\frac r2\right)^{\frac32}\ze(\bx,t)
  =\frac{\sqrt{2\pi}}{K(\frac1{\sqrt2})}
  \sum_{k=0}^\infty4^{-(k+2)}{}_2F_1\left(k+\frac34,\frac12;1;\frac{\rho^2}{r^2}\right)
  \prod_{j=1}^k(4j-1)^2\cdot
    \frac{(ct/r)^{2k}}{(2k)!}\\
  +i\,\frac{K(\frac1{\sqrt2})}{\sqrt{2\pi}}
  \sum_{k=0}^\infty4^{-(k+2)}{}_2F_1\left(k+\frac54,\frac12;1;\frac{\rho^2}{r^2}\right)
   \prod_{j=1}^k(4j+1)^2\cdot
    \frac{(ct/r)^{2k+1}}{(2k+1)!}.
\end{multline}
\end{widetext}
\end{enumerate}

\subsection{Photon position eigenfunctions for general frames}\label{sec.gen}

The method followed in appendix~\ref{app.calc} to derive the explicit formulas for the photon
position eigenfunctions presented in the previous subsection admits several straightforward
generalizations that we shall next discuss.

To begin with, the calculations in appendix~\ref{app.calc} can be performed starting from the
following variant of the standard frame~\eqref{basis}, in which the vector $\bm e_3$ is replaced
by an arbitrary unit vector $\bn$:
\begin{equation}\label{basisn}
  \bm{E}_1(\bk)=\frac{\bk\times (\bk\times\bm{n})}{kk_{\perp}},
  \quad  \bm{E}_2(\bk)=-\frac{\bk\times\bm{n}}{k_{\perp}},  
  \quad \bm{E}_3(\bk)=\frac{\bk}{|k|},
\end{equation}
where
\[
  k_{\perp}=|\bk-(\bn\cdot\bk)\bn| =(k^2-(\bn\cdot\bk)^2)^{\frac12}.
\]
The explicit expressions for the LP and RS eigenfunctions in this more general frame are simply
obtained from eq.~\eqref{Psi0gen} by setting
\begin{subequations}\label{e3ton}
  \begin{equation}\label{cylbasis}
    \bm e_\rho=\frac{\bx_\perp}{x_\perp},\qquad \bm e_\psi=\frac{\bn\times\bx}{x_\perp},\qquad
    \bm e_3=\bn,
  \end{equation}
  where
  \begin{equation}\label{bn}
    \bx_\perp=\bx-(\bn\cdot\bx)\bn,\quad
    x_\perp=|\bx_\perp|=\left(r^2-(\bn\cdot\bx)^2\right)^{\frac12}
  \end{equation}
\end{subequations}
and $\th=\arccos(\bn\cdot\bx/r)$ is the angle between $\bn$ and $\bx$.

Let us next determine the position eigenfunctions corresponding to the general choice of
frame~\eqref{Eiprime}, which differs from the standard one in eq.~\eqref{basis} by a rotation
about $\bE_3$ by a variable angle $\arctan (b(\bk)/a(\bk))$. From eq.~\eqref{crj} and the
definition~\eqref{Eiprime} of the new frame vectors we immediately obtain the formulas
\begin{subequations}\label{bPsipr}
  \begin{align}
    (\hbar
    c)^{-\frac\be2}&\bPsi_{\bm0}^{(1)}(\bx)\notag\\
                   &=\int_{\RR^3}\frac{\dd^3k}{(2\pi)^{3/2}}\big(a(\bk)\bE_1(\bk)
                     -b(\bk)\bE_2(\bk)\big)e^{i\bk\cdot\bx},\label{bPsipr1}\\
     (\hbar
    c)^{-\frac\be2}&\bPsi_{\bm0}^{(2)}(\bx)\notag\\
                   &=\int_{\RR^3}\frac{\dd^3k}{(2\pi)^{3/2}}\big(b(\bk)\bE_1(\bk)
                     +a(\bk)\bE_2(\bk)\big)e^{i\bk\cdot\bx},\label{bPsipr2}
  \end{align}
\end{subequations}
From the calculation in appendix~\ref{app.calc} it is straightforward to obtain the analogues of
eqs.~\eqref{1.3}-\eqref{gradJ}, which we can concisely express in terms of the eigenfunctions with
definite helicity~$\si$,
$\bPsi_{\bm0,\si}=\frac1{\sqrt2}(\bPsi_{\bm0}^{(1)}+i\si\bPsi_{\bm0}^{(2)})$, as
\begin{equation}
  \label{bPsisipr}
  \bPsi_{\bm0,\si}(\bx)=\frac{(\hbar c)^{\frac\be2}}{\sqrt2}
  \left(C_\si^{(1)}(\bx)+\si C_\si^{(2)}(\bx)\right)
\end{equation}
with
\begin{subequations}\label{Csi}
  \begin{align}
    \label{Csi1}
    C_{\si}^{(1)}&=(2\pi)^{-\frac32}\left(\bm e_3\Delta-\nabla\frac{\pd}{\pd
                   x_3}\right)I_{\si,\be}(\bx),\\
    C_{\si}^{(2)}&=(2\pi)^{-\frac32}\bm e_3\times\nabla I_{\si,\be+2}(\bx),
                   \label{Csi2}
  \end{align}
\end{subequations}
where the integral $I_{\si,\be}$ is defined by the analogue of eq.~\eqref{1.2}:
\begin{equation}\label{Isibe}
  I_{\si,\be}=\int_{\RR^3}\frac{\dd^3k}{k_\perp}\,
  k^{\frac\be2-1}\big(a(\bk)+i\si b(\bk)\big)e^{i\bk\cdot\bx}.
\end{equation}
In fact, from the real character of the functions $a(\bk)$ and $b(\bk)$ it follows that
\[
  I_{-,\be}(\bx)=I_{+,\be}(-\bx)^*,
\]
and therefore
\begin{equation}\label{bPsipm}
  \bPsi_{\bm0,-}(\bx)=\bPsi_{\bm0,+}(-\bx)^*.
\end{equation}
Hence we only need compute (for instance) $I_{+,\be}$ and $\bPsi_{\bm0,+}$.

Consider, for example, the frame determined by the choice
\[
  a(\bk)=\frac{k_1}{k_\perp},\qquad b(\bk)=-\frac{k_2}{k_\perp},
\]
as in ref.~\cite{DPTT21}. In this case the integral $I_{+,\be}$ can be easily performed in
cylindrical coordinates ($k_\perp,\phi,k_3$) in $\bk$ space, with the result
\begin{widetext}
  \begin{equation*}
    I_{+,\be}
    =\int_0^\infty\dd k_\perp\int_{-\infty}^\infty \dd k_3
      \frac{e^{ik_3x_3}}{(k_3^2+ k_\perp^2)^{\frac12-\frac\be4}}\int_0^{2\pi}
      \dd\phi\, e^{i(\rho k_\perp\cos(\phi-\psi)-i\phi)}
    =2\pi ie^{-i\psi}\int_0^\infty\dd k_\perp\,J_1(\rho k_\perp)
      \int_{-\infty}^\infty
      \frac{e^{ik_3x_3}\dd k_3}{(k_3^2+ k_\perp^2)^{\frac12-\frac\be4}},
  \end{equation*}
\end{widetext}
where $(\rho,\psi,x_3)$ are the cylindrical coordinates of $\bx$. Although this integral can be
evaluated in terms of elementary functions for arbitrary $\be$, we shall focus here on the cases
$\be=0,2$ needed to compute the LP position eigenfunctions. We have
\begin{align*}
  I_{+,0}&=4\pi ie^{-i\psi}\int_0^\infty\dd k_\perp\,J_1(\rho k_\perp)K_0(|x_3|k_\perp)
  =-\frac{4\pi
           i}{\rho}\,e^{-i\psi}\ln s,\\
  I_{+,2}&=4\pi^2 i\,e^{-i\psi}\de(x_3)\int_0^\infty\dd k_\perp\,J_1(\rho k_\perp)
           =\frac{4\pi^2 i}{\rho}e^{-i\psi}\de(x_3),
\end{align*}
with $s=|\cos\th|$. From the previous formulas and eqs.~\eqref{bPsisipr}-\eqref{Csi}, after a
straightforward calculation we obtain the following explicit expression for the LP position
eigenfunction with helicity $+1$:
\begin{subequations}\label{bPsiLPD}
  \begin{multline}
    \label{bPsiLPDp}
    \bPsi_{\bm0,+}(\bx)=\frac{e^{-i\psi}}{\sqrt\pi\,r^4}\left\{\left[i\,\frac{x_3^2-\rho^2}{x_3}
        -\pi\rho^2\de(x_3)\right]\bm e_\rho\right.\\
    \left.+r^2\left[\frac{1}{x_3}-i\pi\de(x_3)\right]\bm e_\psi
      -2i \rho\bm e_3\right\}.
  \end{multline}
  Its counterpart with helicity $-1$ is easily found from eq.~\eqref{bPsipm}:
  \begin{multline}
    \label{bPsiLPDm}
    \bPsi_{\bm0,-}(\bx)=\frac{e^{i\psi}}{\sqrt\pi\,r^4}\left\{\left[i\,\frac{x_3^2-\rho^2}{x_3}
        -\pi r^2\de(x_3)\right]\bm e_\rho\right.\\
    \left.-r^2\left[\frac{1}{x_3}+i\pi\de(x_3)\right]\bm e_\psi
      +2i\rho\bm e_3\right\}.
  \end{multline}
\end{subequations}
As mentioned above, a slightly more general version of these formulas is obtained by replacing the
$\bm e_3$ vector by a generic unit normal vector $\bn$ (with $\psi$ the azimuthal angle about
$\bn$, measured in the anticlockwise direction), and using eqs.~\eqref{e3ton}.
\section{Analysis of the singularities}\label{sec.sing}

We shall next discuss in detail the singularities away from the origin of the LP and RS photon
position eigenfunctions. To this end, we recall the well-known relations
\begin{subequations}\label{KEser0}
  \begin{align}
    K(\ka)&=\frac\pi2\left(1+\frac{\ka^2}4+O(\ka^4)\right),\label{Kser0}\\
    E(\ka)&=\frac\pi2\left(1-\frac{\ka^2}4+O(\ka^4)\right),\label{Eser0}
  \end{align}
\end{subequations}
valid for $\ka\to0+$, together with
\begin{subequations}\label{KEser1}
  \begin{align}
    K\bigl(\sqrt{1-\vep^2}\,\bigr)&=-\ln\vep+O(1),\label{Kser1}\\
    E\bigl(\sqrt{1-\vep^2}\,\bigr)&=1+O(\vep^2\ln\vep),\label{Eser1}
  \end{align}
\end{subequations}
valid for $\vep\to0+$.

Consider, first, the LP eigenfunction $\bPsi_{\bm0,\si}$ in eq.~\eqref{Psi0gen} with $\be=0$, with
components $P_j$ given by eqs.~\eqref{LPPs}. The apparent singularities of this function away from
the origin are the horizontal plane $\th=\pi/2$ and the $x_3$ axis $\th=0,\pi$. The latter,
however, is \emph{not} a true singularity. Indeed, from eqs.~\eqref{Kser0}-\eqref{Eser0} it
follows that as $\th\to0+$ or $\th\to\pi-$ we have
\begin{align*}
  2&E(\sin\th)\cot(2\th)-K(\sin\th)\cot\th\\
   &=\frac\pi2(\cot\th-\tan\th)\left(1-\frac14\sin^2\th+O(\sin^4\th)\right)\\
   &\qquad-\frac\pi2\cot\th\left(1+\frac14\sin^2\th+O(\sin^4\th)\right)\\
  &=\frac{3\pi}4\sgn\bigl(\th-\tfrac\pi2\bigr)\sin\th+O(\sin^3\th)
\end{align*}
and
\begin{align*}
  K(\sin\th)-2E(\sin\th)=-\frac\pi2+O(\sin^2\th).
\end{align*}
Therefore near the $x_3$ axis (i.e., near $\th=0$ or $\th=\pi$) we have
\begin{align*}
  \bPsi_{\bm0,\si}(\bx)&=\frac{\sqrt\pi}4\,r^{-3}
                         \left(3\sgn\bigl(\th-\tfrac\pi2\bigr)\sin\th\,
                         \bm e_\rho-2\bm e_z+O(\sin^2\th)\right)\\
                       &=-\frac{\sqrt{\pi}}2\,r^{-3}\bm e_z+O(\sin\th).
\end{align*}

On the other hand, we shall next show that $P_\rho$ and $P_z$ are singular on the horizontal plane
$\th=\pi/2$. Indeed, eqs.~\eqref{Kser1}-\eqref{Eser1} imply that as $\th\to\pi/2$ we have
\begin{align*}
  \sqrt{\pi}\,P_\rho
  &=2E(\sin\th)\cot(2\th)-K(\sin\th)\cot\th\\
  &=(\cot\th-\tan\th)\left(1+O(\cos^2\th\ln\cos\th)\right)\\
  &\qquad-\cot\th\left(-\ln\cos\th+O(1)\vphantom{{}^2}\right)\\
  &=
    -\tan\th+O(\cos\th\ln\cos\th)\\
  &=-\sec\th+O(\cos\th\ln\cos\th),\\
  \sqrt{\pi}\,P_z&=K(\sin\th)-2E(\sin\th)=-\ln\cos\th+O(1),
\end{align*}
From these equations we obtain the following formula for the behavior of the LP position
eigenfunctions with well-defined helicity~$\si$ near the horizontal plane $x_3=0$:
\begin{multline}
  \label{Psisias}
  \bPsi_{\bm0,\si}(\bx)=-\pi^{-\frac12}\,r^{-3}\bigg(\sec\th\,\bm e_\rho
  +\si\pi\de\bigl(\th-\tfrac\pi2\bigr)\bm e_\psi+\ln\cos\th\bm e_z\\
  +O(\cos\th\ln\cos\th)\bigg).
\end{multline}

Consider, next, the RS eigenfunction $\cF_{\bm0},\si$ given by eqs.~\eqref{RSPs}. Again, the
infinite singularity of this eigenfunction on the $x_3$ axis is only apparent, since for
$s=|\cos\th|\to1-$ we have $\ka_1\to0+$, and therefore (omitting, for the sake of conciseness, the
module of the elliptic integrals)
\begin{align*}
  A_1K+B_1E&=\frac{\pi}2\big(A_1(1)+B_1(1)\big)+O(1-s)\\&=O(1-s)
  \implies
  \frac{A_1K+B_1E}{\sqrt{1-s}}=O(\sqrt{1-s}\,).
  \vrule width0pt depth12pt
\end{align*}
Thus $P_\rho$ in eq.~\eqref{RSPrho} vanishes near the $x_3$ axis (at least) as
$\sqrt{1-|\cos\th|}$. On the other hand, the function $P_z$ behaves near this axis as
\begin{equation*}
  \frac{\pi}{16K(\frac1{\sqrt2})}
  (A_3(1)+B_3(1))
  =-\frac{3\pi}{8K(\frac1{\sqrt2})},
\end{equation*}
and is therefore non-singular. Let us analyze, finally, the behavior of the function $P_\psi$ in
eq.~\eqref{RSPpsi} near the $x_3$ axis. In this case $\ka_1=O(1-s)$ as $s\to1-$, and therefore
\[
  K(\ka_1)=\frac\pi2+O((1-s)^2),\quad E(\ka_1)=\frac\pi2+O((1-s)^2)
\]
by eqs.~\eqref{Kser0}-\eqref{Eser0}. On the other hand, from eq.~\eqref{ABtAtB1} we have
\begin{align*}
  A_2&=2-\frac{19}2(1-s)+O((1-s)^2),\\ B_2&=-2+12(1-s)+O((1-s)^2)
\end{align*}
as $s\to1-$. Hence
\[
  \frac{A_2K(\ka_1)+B_2E(\ka_1)}{s^{\frac{3}{2}}(1-s)^{\frac{1}{2}}}
  =\frac{5\pi}{4}\,(1-s)^{\frac12}+O\bigl((1-s)^{3/2}\bigr)
\]
tends to zero as $s\to1-$.

We shall next determine the behavior of the RS eigenfunction $\cF_{\bm0.\si}$ near the horizontal
plane $\th=\pi/2$, or equivalently $s=0$. It suffices to note that as $\th\to\frac\pi2$ the
modulus $\ka_1$ of the elliptic integrals tends to $1/\sqrt2$, so that
\begin{align*}
  A_1K+B_1E&=2K\bigl(\tfrac1{\sqrt2}\bigr)+O(s^{\frac{1}{2}}),\\
  A_3K+B_3E&=-4s K\bigl(\tfrac1{\sqrt2}\bigr)
                           +O(s^{\frac{3}{2}}).
\end{align*}
Therefore near the horizontal plane we have
\begin{align*}
  P_\rho(\th)
  &=
   \sgn\bigl(\th-\tfrac\pi2\bigr)(2s)^{-\frac{3}{2}}\big(1+O(s^{\frac12})\big),\\
  P_z(\th)&=-(2s)^{-\frac{1}{2}}(1+O(s^{\frac12})\big),
\end{align*}
which both diverge as $s\to0+$. On the other hand, when $s\to0+$ we have
\[
  A_2K+B_2E=4E\bigl(\tfrac1{\sqrt2}\bigr)-2K\bigl(\tfrac1{\sqrt2}\bigr)+O(s^{\frac{1}{2}}).
\]
From the identity
\[
  E\bigl(\tfrac1{\sqrt2}\bigr)=\frac{\pi}{2K(\frac1{\sqrt2})}+\frac12K\bigl(\tfrac1{\sqrt2}\bigr)
\]
we then obtain
\[
  A_2K+B_2E=\frac{2\pi}{K(\frac1{\sqrt2})}+O(s^{\frac{1}{2}}),
\]
and therefore
\[
  P_\psi(\th)=\frac{s^{-\frac{3}{2}}}{\sqrt2}\left( 1+O(s^{\frac{1}{2}})\right).
\]
Combining this equation with the analogous ones for $P_\rho$ and $p_z$ derived above we finally
obtain the following formula for the asymptotic behavior near the singular plane $x_3=0$ of the RS
position eigenfunction $\cF_{\bm0,\si}$ with well-defined helicity $\si$:
\begin{multline}
  \label{Fsiasy}
  \cF_{\bm0,\si}(\bx)=\sqrt{\frac{\hbar c}2}\,r^{-\frac72}s^{-\frac32}
  \bigg(\tfrac12\sgn\bigl(\th-\tfrac\pi2\bigr)\bm e_\rho +\si\,\bm e_\psi-s\bm e_z\bigg)
  \\
  \times\left(1+O(s^{\frac12})\right).
\end{multline}

\section{Conclusions}

In this paper we derive closed-form expressions in configuration space for the eigenfunctions with
definite helicity of the Hawton photon position operators. We consider a general class of photon
wave functions, with special emphasis in the Riemann-Silberstein (RS) and the Landau-Peierls (LP)
wave functions. We analyze the position operators associated with general frames in 3D Euclidean
momentum space and, in particular, with the spherical coordinate system (i.e., the standard
frame). Our method is based on calculating the corresponding Fourier integrals using several
identities involving the complete elliptic integrals $K$ and $E$ expressed in terms of the Gauss
hypergeometric function ${}_2F_1$. We are able to provide representations of the position
eigenfunctions in configuration space as products of a simple scalar radial factor by a vector
field expressed as a linear combination in the orthonormal basis
$\{\bm e_\rho,\bm e_\psi,\bm e_z=\bm e_3\}$ of the cylindrical coordinate system. The coefficients
of this vector field only depend on the polar angle $\theta$ trough the complete elliptic
integrals, with the exception of the component $\bm e_\psi$ of the LP eigenfunction, which
simplifies to a single delta function centered at $\theta=\pi/2$. For the standard frame, the LP
and RS eigenfunctions are real, and their radial factors are proportional to $r^{-3}$ and
$r^{-7/2}$, respectively. We also obtain a simple expression for the Hertz potential of the RS
field. Our expressions for the photon position eigenfunctions substantially extend and simplify
those available in, e.g., refs.~\cite{BM17,DPTT21}. These expressions are well suited for a
complete analysis of the singularities of these eigenfunctions, and for deriving simple asymptotic
expansions thereof. In this way we deduce an important generic property of these quantum states,
namely the presence of singular planes in configuration space. This property was already noticed
in some particular cases in the latter references. To understand its origin, we note that these
quantum states are inverse Fourier transforms of functions of $\bk$ which are linear combinations
of tangent vector fields on spheres $S^2$ of variable radius $k$. According to the ``hairy-ball
theorem'', any such tangent vector field should present singularities at some points $\bk_0$ in
momentum space. As a consequence, the inverse Fourier transforms of the photon position
eigenfunctions inherits singular contributions on the corresponding planes perpendicular to
$\bk_0$ in configuration space. For example, in the case of the position operator associated to
the standard polarization vectors $ \bm{E}_1(\bk)$ and $ \bm{E}_2(\bk)$ the singularity
$\theta=\pi/2$ ($z=0$) arises from the singularities on the $k_3$ axis of the eigenfunctions in
momentum space.

\section*{Acknowledgments}
This work was partially supported by grants PID2024-155527NB-I00 from Spain's Mi\-nis\-te\-rio de
Ciencia, Innovación y Universidades and~PR12/24-31565 from Universidad Complutense de Madrid. LMA
would also like to thank E. Olmedilla for numerous helpful discussions.

\appendix
\section{some useful identities involving special functions}\label{app.ids}

Gauss's hypergeometric function ${}_2F_1$ is defined by
\begin{equation}\label{2F1}
  {}_2F_1(a,b;c;z)=\sum_{j=0}^\infty\frac{(a)_j(b)_j}{(c)_j}\,\frac{z^j}{j!},
\end{equation}
where $(t)_j=t(t+1)\cdots(t+j-1)$ is Pochhammer's forward symbol. Here $a$, $b$, and $c$ are
complex numbers, with $c$ not equal to a non-negative integer. Although this series has radius of
convergence $1$, the function defined by it can be analytically continued to the complex $z$ plane
with a branch cut from $1$ to $+\infty$. The derivative of the hypergeometric function with
respect to its argument $z$ follows easily from its definition:
\begin{equation}
  \label{2F1der}
  \frac{\dd}{\dd z}\,{}_2F_1(a,b;c;z)=\frac{ab}c\,{}_2F_1(a+1,b+1;c+1;z)
\end{equation}

The complete elliptic integrals of the first and second kind with modulus $\ka\in[0,1]$ are
respectively defined by
\begin{align*}
  K(\ka)&:=\int_0^{\pi/2}\frac{\mathrm{d}\theta}{\sqrt{1-\ka^2\sin^2\theta}},\\
  E(\ka)&:=\int_0^{\pi/2}\mathrm{d}\theta\,\sqrt{1-\ka^2\sin^2\theta}.
\end{align*}
The derivatives of these functions with respect to the modulus $\ka$ are given by
  \begin{align}
    K'(\ka)&=\frac1{\ka(1-\ka^2)}\,\Big(E(\ka)-(1-\ka^2)K(\ka)\Big),\label{Kp}\\
    E'(\ka)&=\frac1\ka\Big(E(\ka)-K(\ka)\Big)
             \label{Ep}
\end{align}
(cf.~\cite[eq.~(19.4.1)]{OLBC10}. According to~\cite[eqs.~15.9.24-25]{OLBC10}, the elliptic
integrals $K$ and $E$ can be expressed in terms of Gauss's hypergeometric function ${}_2F_1$ as
follows:
\begin{equation}\label{EK2F1}
  K(\ka)=\frac{\pi}2\,{}_2F_1\bigl(\tfrac12,\tfrac12;1;\ka^2\bigr),\quad
  E(\ka)=\frac{\pi}2\,{}_2F_1\bigl(-\tfrac12,\tfrac12;1;\ka^2\bigr).
\end{equation}
In the body of the paper we have also made use of the special value
\begin{equation}\label{GaK}
  \Gamma\!\left(\tfrac14\right)^2=4\sqrt\pi\,K\bigl(\tfrac1{\sqrt2}\bigr).
\end{equation}

We shall denote by $J_\mu(z)$ and $K_\nu(z)$ the Bessel function of the first kind and the
modified Bessel function of the second kind of orders $\mu$ and $\nu$, respectively.
From~\cite[eqs.~(10.22.41) and (10.9.2)]{OLBC10} we have:
\begin{equation}\label{A3}
  \int_0^{\infty} \dd k\, J_0(k)=1,
  \qquad  \int_0^{2\pi} \dd\theta\,e^{iz\cos\theta}= 2\pi\,J_0(z).
\end{equation}
Likewise, from \cite[eq.~10.32.11]{OLBC10} it follows at once that
\begin{equation}
  \label{K0intgen}
  \int_{-\infty}^\infty\dd s\,\frac{e^{ixs}}{(s^2+a^2)^{\nu+\frac12}}
  =\frac{2\sqrt\pi\,|x|^\nu}{(2a)^\nu\Ga(\nu+\frac12)}\,K_\nu(a|x|),
\end{equation}
where $a>0$, $x\ne0$, and $\nu>-1/2$.

The following integral can be computed in terms of Gauss's hypergeometric function:
\begin{multline}\label{JKintgen}
  \int_0^{\infty}\dd s\,s^\nu
  K_\nu(as)J_0(bs)\\=2^{\nu-1}a^{-\nu-1}\sqrt\pi\,\Ga\biggl(\nu+\frac12\biggr)
  {}_2F_1\biggl(\nu+\frac12,\frac12;1;-\frac{b^2}{a^2}\biggr),
\end{multline}
where $\nu\ge0$ and $a>0$ (cf.~\cite[eq.~10.43.26]{OLBC10}). Using the linear
transformation~\cite[eq.~15.8.1]{OLBC10}
\[
  {}_2F_1(\al,\be;\ga;-z)=(1+z)^{-\al}{}_2F_1\left(\al,\ga-\be;\ga;\frac{z}{1+z}\right),
\]
the previous equation can be expressed in the more convenient form
\begin{multline}\label{JKintgen2}
  \int_0^{\infty}\dd s\,s^\nu
  K_\nu(as)J_0(bs)=\frac{2^{\nu-1}\sqrt\pi\,a^\nu}{(a^2+b^2)^{\nu+\frac12}}\,
  \Ga\bigl(\nu+\tfrac12\bigr)\\
  \times{}_2F_1\biggl(\nu+\frac12,\frac12;1;\frac{b^2}{a^2+b^2}\biggr),
\end{multline}
In particular, when $\nu=0$ it follows from eq.~\eqref{EK2F1} that
\begin{equation}
  \label{JKintbe0}
  \int_0^{\infty}\dd s\,
  K_0(as)J_0(bs)=(a^2+b^2)^{-\frac12}K\left(\frac{b}{\sqrt{a^2+b^2}}\right)\,.
\end{equation}

The quartic transformation formulas among hypergeometric functions
\begin{widetext}
  \begin{align} {}_2F_1\bigl(\tfrac34,\tfrac12;1;1-z^4\bigr)
    &=z^{-1}\sqrt{\frac2{1+z^2}}\,
    {}_2F_1\bigl(\tfrac12,\tfrac12;1;m(z)\bigr)
\label{2F1tr1}\\
    {}_2F_1\bigl(\tfrac54,\tfrac12; 1; 1-z^4\bigr)
 &=z^{-3}\sqrt{\frac2{1+z^2}}\,\Big[2(1+z^2)\,
  {}_2F_1\bigl(-\tfrac12,\tfrac12;1;m(z)\bigr)
 -(z^2+z+1)\,{}_2F_1\bigl(\tfrac12,\tfrac12;1;m(z)\bigr)\Big],
\label{2F1tr2}
  \end{align}
\end{widetext}
with
\begin{equation}
  \label{mRS2}
  m(z)=\frac{(1-z)^2}{2(1+z^2)},
\end{equation}
are used in the computation of the RS eigenfunctions~$\cF_{\bm0}^{(i)}$
(cf.~appendix~\ref{app.calc}). To prove these formulas, we start by establishing the identity
\begin{equation}
  \label{2F1id}
  {}_2F_1\bigl(\tfrac14,\tfrac12;1;1-z^4\bigr)=\sqrt{\frac{2}{1+z^2}}\,
  {}_2F_1\bigl(\tfrac12,\tfrac12;1;m(z)\bigr).
\end{equation}
To this end, we first apply to the left-hand side the transformation~\cite[eq.~15.8.21]{OLBC10}
\begin{multline*}
  {}_2F_1\biggl(a,b;2b;\frac{4x}{(1+x)^2}\biggr)\\
  =(1+x)^{2a}{}_2F_1\biggl(a,a-b+\frac12;b+\frac12;x^2\biggr),
\end{multline*}
with $a=1/4$, $b=1/2$, and (if $0\le x\le 1$)
\[
  \frac{4x}{(1+x)^2}=1-z^4\iff x=\frac{1-z^2}{1+z^2},
\]
to obtain
\begin{equation*}
  {}_2F_1\biggl(\frac14,\frac12;1;1-z^4\biggr)
  =\sqrt{\frac2{1+z^2}}\,
  {}_2F_1\biggl(\frac14,\frac14;1;\left(\frac{1-z^2}{1+z^2}\right)^{\!\!2}\,\biggr).
\end{equation*}
We next apply to the right-hand side of the last identity the
transformation~\cite[eq.~15.8.18]{OLBC10}
\begin{multline*}
  {}_2F_1\left(a, b; a+b+\frac{1}{2}; 4y(1-y)\right)\\
  = {}_2F_1\left(2a, 2b; a+b+\frac{1}{2}; y\right)
\end{multline*}
with $a=b=1/4$, and
\begin{equation}\label{yeq}
  4y(1-y)=\left(\frac{1-z^2}{1+z^2}\right)^{\!\!2},
\end{equation}
obtaining (for $0\le y\le 1/2$)
\[
    {}_2F_1\bigl(\tfrac14,\tfrac12;1;1-z^4\bigr)=\sqrt{\frac{2}{1+z^2}}\;
  {}_2F_1\bigl(\tfrac12,\tfrac12;1;y\bigr).
\]
Solving for $y$ in eq.~\eqref{yeq} then yields
\[
  y=\frac{(1-z)^2}{2(1+z^2)}=m(z),
\]
thus establishing eq.~\eqref{2F1id}.

Equations~\eqref{2F1tr1}-\eqref{2F1tr2} can be easily derived from the identity~\eqref{2F1id} just
proved as follows. To begin with, applying the standard Euler
transformation~\cite[eq.~15.8.1]{OLBC10}
\[
  {}_2F_1(a,b;c;u)=(1-u)^{c-a-b}{}_2F_1(c-a,c-b;c;u)
\]
to the left-hand side of eq.~\eqref{2F1id} we immediately obtain eq.~\eqref{2F1tr1}. Likewise, to
derive eq.~\eqref{2F1tr2} it suffices to apply to eq.~\eqref{2F1id} the raising transformation
\begin{equation}\label{2F1ap1}
  \frac1a\left(u\frac{\dd}{\dd u}+a\right){}_2F_1(a,b;c;u)={}_2F_1(a+1,b;c;u)
\end{equation}
with $u=1-z^4$ and
\[
  \frac{\dd}{\dd u}=-\frac{1}{4z^3}\,\frac{\dd}{\dd z}.
\]

\section{a simple proof of eqs.~\eqref{Qbetaall}}\label{app.HPO}

In this appendix we provide a simple proof of eqs.~\eqref{Qbetaall} for the photon position
operator~\eqref{RSQj}. Obviously, it suffices to prove the corresponding equations~\eqref{H2all},
from which~\eqref{Qbetaall} follow from the elementary relation
\begin{align*}
  i\,k^{\be/2}\,\mathbb E\,\frac{\partial}{\partial k_j}\,\mathbb E^{-1}k^{-\be/2}
  &=i\,\mathbb E\,\frac{\partial}{\partial k_j}\,\mathbb E^{-1}
    +i\,k^{\be/2}\frac{\partial}{\partial k_j}k^{-\be/2}\\
  &=i\,\mathbb E\,\frac{\partial}{\partial k_j}\,\mathbb E^{-1}-\frac{i\be\bk}{2k^2}.
\end{align*}

To prove eqs.~\eqref{H2all}, we start by rewriting eq.~\eqref{po} as
\[
  \frac1i\,Q_j=\frac{\pd}{\pd k_j}+\EE\,\frac{\pd\EE^{-1}}{\pd k_j}=:\frac{\pd}{\pd k_j}+M_j.
\]
To simplify the matrix operator $M_j$, note that by definition
\[
  E_{ij}=\frac1{h_j}\frac{\pd k_i}{\pd q_j},
\]
where $\bq=(\th,\phi,k)$ are spherical coordinates in momentum space and
\[
  h_j=\left\vert\pdf{\bk}{q_j}\right\vert\implies (h_1,h_2,h_3)=(r,r\sin\th,1).
\]
We shall also use the elementary identities
\[
  (\EE^{-1})_{ij}=E_{ji}=h_i\pdf{q_i}{k_j}
\]
and
\[
  \pdf{}{k_j}=\sum_{i=1}^3\pdf{q_i}{k_j}\pdf{}{q_i}=\sum_{i=1}^3\frac{E_{ji}}{h_i}\pdf{}{q_i}.
\]
From the previous relations it follows that
\begin{align*}
  (M_j)_{il}
  &=\sum_{m,n=1}^3E_{im}\frac{E_{jn}}{h_n}\pdf{E_{lm}}{q_n}=
    \sum_{n=1}^3\frac{E_{jn}}{h_n}\sum_{m=1}^3E_{im}\pdf{E_{lm}}{q_n}\\
    &=-\sum_{n=1}^3\frac{E_{jn}}{h_n}\sum_{m=1}^3\pdf{E_{im}}{q_n}E_{lm}
  =-\sum_{n=1}^3\frac{E_{jn}}{h_n}\left(\pdf{\EE}{q_n}\,\EE^\T\right)_{il},
\end{align*}
where in the third equality we have used the identity
\[
  E_{im}\pdf{E_{lm}}{q_n}=-\pdf{E_{im}}{q_n}E_{lm}.
\]
Taking into account that the matrix elements $E_{ij}$ are all independent of $k=q_3$ we can
rewrite the previous equation as
\begin{equation}\label{Mj1}
  M_j=-\frac1k\left(E_{j1}\pdf{\EE}{\th}+\frac{E_{j2}}{\sin\th}\pdf{\EE}{\phi}\right)\EE^\T.
\end{equation}
A direct calculation shows that
\[
  \pdf{\EE}{\th}\,\EE^\T=\frac1i\big(\cos\phi\, S_2-\sin\phi\, S_1\big),\qquad
  \pdf{\EE}{\phi}\,\EE^\T=\frac{S_3}i,
\]
so that eq.~\eqref{Mj1} becomes
\begin{equation}
  \label{Mj2}
  M_j=-\frac1{ik}\left[E_{j1}\big(\cos\phi\, S_2-\sin\phi\, S_1\big)+\frac{E_{j2}}{\sin\th}\,
    S_3\right].
\end{equation}
In particular, since $E_{32}=0$ and $E_{31}=-\sin\th$ we have
\[
  M_3=\frac{1}{ik^2}(k_1S_2-k_2S_1)=\frac{1}{ik^2}(\bk\times\bS)_3.
\]
Likewise, from eq.~\eqref{Mj2} with $j=1$ we obtain
\begin{widetext}
  \begin{align*}
    M_1-\frac{1}{ik^2}(\bk\times\bS)_1
    &=-\frac1{ik}\left[\cos\th\,\cos\phi\big(\cos\phi\,
      S_2-\sin\phi\, S_1\big)
      -\frac{\sin\phi}{\sin\th}S_3\right]-\frac1{ik}\big(\sin\th\,\sin\phi\, S_3-\cos\th\,S_2\big)\\
    &=\frac1{ik}\left[\sin\phi\cos\phi\cos\th\,S_1+\sin^2\phi\cos\th\, S_2
      +\sin\phi\frac{\cos^2\th}{\sin\th}\,S_3\right]=\frac1{ik^2}\sin\phi\cot\th\,\bk\cdot\bS.
  \end{align*}
\end{widetext}
A similar calculation shows that
\[
  M_2-\frac{1}{ik^2}(\bk\times\bS)_2=-\frac1{ik^2}\cos\phi\cot\th\,\bk\cdot\bS.
\]
Since $\bE_2=(-\sin\phi,\cos\phi,0)$, we can rewrite the previous formulas for the matrix
operators $M_{j}$ ($j=1,2,3$) in vector form as
\begin{equation}
  \label{Mvec}
  i\bM=\frac1{k^2}\big(\bk\times\bS-\bE_2\cot\th\,\bk\cdot\bS\big)=
  \frac1{k^2}\,\bk\times\bS-\frac{\cot\th}k\,\bE_2\Si\,,
\end{equation}
which immediately yields eq.~\eqref{H2alt}:
\begin{equation}\label{Qvec}
  \bQ=i\nabla_{\bk}+i\bM=i\nabla_{\bk}+\frac1{k^2}\,\bk\times\bS-\frac{\cot\th}k\,\bE_2\Si\,.
\end{equation}
To obtain eq.~\eqref{H2}, it suffices to note that
\begin{equation}
  \label{SiS3}
  \Si=\EE S_3\EE^{-1}.
\end{equation}
Indeed,
\[
  \Si_{jl}=\frac1{k}\sum_{n=1}^3k_n(S_n)_{jl}=-\frac ik\sum_{n=1}^3\vep_{njl}k_n,
\]
while
\begin{align*}
  \left(\EE S_3\EE^{-1}\right)_{jl}
  &=\sum_{m,n=1}^3E_{jm}(S_3)_{mn}E_{ln}\\
  &=-i\sum_{m,n=1}^3E_{jm}\vep_{3mn}E_{ln}
  =-i\big(E_{j1}E_{l2}-E_{j2}E_{l1}\big)\\
  &       =-i\big[(\bE_{1})_j(\bE_{2})_l-(\bE_{1})_l(\bE_{2})_j\big]\\
    &=-i\sum_{n=1}^3\vep_{njl}(\bE_1\times\bE_2)_n=-i\sum_{n=1}^3\vep_{njl}(\bE_3)_n\\
  &=-\frac{i}k\sum_{n=1}^3\vep_{njl}k_n.
\end{align*}
Note, in conclusion, that if we use the general orthonormal frame~\eqref{basisn} defined in
appendix~\ref{app.calc} equation~\eqref{QbetaSi} for the position operator becomes
\begin{equation}
  \label{Qvecn}
  \bQ=i\nabla_{\bk}-\frac{i\be}{2k^2}\bk+\frac1{k^2}\,\bk\times\bS
  +\frac{\cot\th}{kk_\perp}\,\bk\times\bn\,\Si\,,
\end{equation}
where $k_\perp=[k^2-(\bk\cdot\bn)^2]^{1/2}$ and $\th$ is the angle between $\bk$ and the constant
unit vector $\bn$.

\section{normalization of the position eigenfunctions in configuration space}\label{app.norm}

Applying Plancherel's theorem we have
\begin{align*}
  \big(\,\bPsi_{\bq}^{(j)},\bPsi_{\bq'}^{(j')}\big)
  &=\int_{\RR^3}\dd^3k\,\big(\bpsi_{\bq}^{(j)}\big)^*\big(\bpsi_{\bq'}^{(j')}\big)\\
  &=\int_{\RR^3}\dd^3(\hbar ck)^\be
    \bE_j(\bk)\cdot\bE_{j'}(\bk)e^{i(\bq-\bq')\bk}\\
  &=\de_{jj'}\int_{\RR^3}\dd^3(\hbar ck)^\be e^{i(\bq-\bq')\bk}.
\end{align*}
When $\be=0$ the last integral equals $(2\pi)^3\de(\bq-\bq')$, from which eqs.~\eqref{norm0}
immediately follow. On the other hand, when $\be\ne0$ the integral can be evaluated in spherical
coordinates, taking the $k_3$ axis in the direction of the vector $\bp:=\bq-\bq'$:
\begin{align*}
  \int_{\RR^3}\dd^3k\,k^\be e^{i\bp\cdot\bk}
  &=2\pi\int_0^\infty\dd k\,k^{\be+2}\int_0^\pi
    \dd\th\sin\th\,e^{ipk\cos\th}\\
  &=\frac{4\pi}p\int_0^\infty\dd k\, k^{\be+1}\sin(pk)\\
  &=\frac{4\pi}{p^{\be+3}}\int_0^\infty\dd s\, s^{\be+1}\sin s.\\
\end{align*}
Although the last integral converges only for $-3<\be<-1$, for $\be\ge-1$ we regularize it as
\begin{multline*}
  \lim_{\vep\to0+}\Im\int_0^{\infty}\dd s\, s^{\be+1}e^{is}e^{-\vep s}\\
  =\lim_{\vep\to0+}\Im\left[(\vep-i)^{-(\be+2)}\int_0^{(\vep-i)\infty}\dd t\, t^{\be+1}e^{-t}\right].
\end{multline*}
Rotating the integration line back to the positive real axis and letting $\vep\to0+$ we 
obtain
\begin{align*}
  \lim_{\vep\to0+}\Im\int_0^{\infty}\dd s\, s^{\be+1}e^{is}e^{-\vep s}
  &=\Im\left[(-i)^{-(\be+2)}\Ga(\be+2)\right]\\
  &=-\sin\left(\tfrac{\pi\be}2\right)\Ga(\be+2).
\end{align*}
Putting everything together we finally obtain eqs.~\eqref{norm1}.
 
\section{calculation of the position eigenfunctions in configuration space}
\label{app.calc}
\subsection{The function \texorpdfstring{$\bPsi_{\bm 0}^{(1)}(\bx)$}{\relax}}
\label{sec:Psi1}
\smallskip\noindent
From the expression of $\bm{E}_1$ in \eqref{bE1} and eq.~\eqref{crj} we may write
\begin{align}
  (\hbar c)^{-\frac\be2}\bPsi_{\bm{0}}^{(1)}(\bx)
  &=(2\pi)^{-\frac32}\mkern-6mu\int_{\mathbb{R}^3}
  \dd^3k\,\big(\bk\times (\bk\times \bm{e}_3)\big)\frac{k^{\frac\be2-1}}{k_{\perp}}
  e^{i \bk \cdot \bx }\notag\\
  &=
  -(2\pi)^{-\frac32}\bm{\nabla}\times
  \Big( \bm{\nabla}\times \big(I_\be(\bx)\bm{e}_3\big)\Big),
  \label{1.1}
\end{align}
where
\begin{equation}\label{1.2}
  I_\be(\bx):=\int_{\mathbb{R}^3} \dd^3k\, \frac{k^{\frac\be2-1}}{k_{\perp}}\,
  e^{i \bk \cdot \bx }.
\end{equation}
Now, from the identity
$\bm{\nabla}\times (\bm{\nabla}\times \bm{A})=\bm{\nabla}(\bm{\nabla}\cdot
\bm{A})-\triangle\bm{A}$ it follows that
\begin{equation}\label{1.3}
  \bPsi_{\bm{0}}^{(1)}(\bx)
  =(\hbar c)^{\frac\be2}(2\pi)^{-\frac32}\Big((\triangle I_\be) {\bm e}_3
  -\bm{\nabla}\frac{\partial I_\be}{\partial x_3}\Big).
\end{equation}
Moreover, applying the identity \eqref{K0intgen} with $\nu=-\be/4$ we obtain
\begin{align}
  I_\be(\bx)
  &= \int_{\mathbb{R}^2} \dd^2k_{\perp}\,
    \frac{ e^{i \bk_{\perp} \cdot \bx_{\perp} }}{k_{\perp}}
    \int_{-\infty}^{\infty} \dd k_3\, \frac{ e^{i k_3  x_3
    }}{(k_{\perp}^2+k_3^2)^{\frac12-\frac\be4}}\notag\\
  &=
    \frac{2^{\frac\be4+1}\sqrt\pi}{\Ga\left(\frac12-\frac\be4\right)}
    |x_3|^{-\frac\be4}\int_{\mathbb{R}^2}
    \frac{\dd^2k_{\perp}}{k_\perp}\,
    e^{i \bk_{\perp} \cdot \bx_{\perp} }\, k_\perp^{\frac\be4}K_{\frac\be4}(k_{\perp} |x_3|),
    \label{Ibeta}
\end{align}
where $ \bk_{\perp }=(k_1,k_2),\, \bx_{\perp} =(x_1,x_2)$ and $K_\nu$ is the modified Bessel
function of the second kind and order $\nu$. Note that for $k_\perp\ne0$ the integral over $k_3$
in the previous equation converges absolutely for $\be<0$ and conditionally for $0\le\be<2$.
Expressing the last integral for $I_\be$ in polar coordinates $(k_{\perp}, \phi)$ and
$(\rho, \psi)$ for $\bk_{\perp}$ and $\bx_{\perp}$, respectively, and using the
identities~\eqref{A3} and~\eqref{JKintgen2}, we obtain
\begin{widetext}
\begin{align}
  I_\be(\bx)
  &= \frac{2^{\frac\be4+1}\sqrt\pi}{\Ga\left(\frac12-\frac\be4\right)}|x_3|^{-\frac\be4}
    \int_0^{\infty} \dd k_\perp\, k_\perp^{\frac\be4}  K_{\frac\be4}(|x_3|k_\perp)
    \int_0^{2\pi} \dd\phi \,e^{i \rho k_\perp \cos(\phi-\psi)}\notag\\
  &=\frac{2^{\frac\be4+2}\pi^{\frac32}}{\Ga\left(\frac12-\frac\be4\right)}|x_3|^{-\frac\be4}
    \int_0^{\infty} \dd  k_\perp\,  k_\perp^{\frac\be4}K_{\frac\be4}(|x_3| k_\perp)\,J_0(\rho k_\perp)
  =\pi^2\frac{\Ga(\frac12+\frac\be4)}{\Ga(\frac12-\frac\be4)}\,\left(\frac2r\right)^{\frac\be2+1}
  {}_2F_1\left(\frac12+\frac\be4,\frac12;1;\sin^2\th\right),
  \label{main}
\end{align}
\end{widetext}
where $r=|\bx|$, $\th$ is the angle between the vector $\bx$ and the $\bm e_3$ axis, and $J_0$ is
the Bessel function of the first kind and order zero. In the two main cases of interest, $\be=0$
and $\be=1$, the function $I_\be$ can be expressed in terms of the complete elliptic integral of
the first kind $K(\ka)$. Indeed, for $\be=0$ eqs.~\eqref{main} and~\eqref{EK2F1} immediately yield
\begin{equation}
  I_0(\bx)=\frac{4\pi}r\,K(\sin\th).
  \label{Ibe0}
\end{equation}
Similarly,  from eq.~\eqref{main} with $\be=1$ we obtain
\[
  I_1(\bx)=\pi^2\,\frac{\Ga(\frac34)}{\Ga(\frac14)}\,\left(\frac2r\right)^{\frac32}
  {}_2F_1\biggl(\frac34,\frac12;1;\sin^2\th\biggr).
\]
The quotient of gamma functions can be simplified using the reflection identity for the gamma
function
\begin{equation}\label{gammar}
  \Ga(z)\Ga(1-z)=\frac{\pi}{\sin(\pi z)}
\end{equation}
together with eq.~\eqref{GaK}, namely
\[
  \frac{\Ga(\frac34)}{\Ga(\frac14)}=\frac{\sqrt2\,\pi}{\Ga(\frac14)^2}
  =\frac{\sqrt{2\pi}}{4K(\frac1{\sqrt2})}.
\]
We thus obtain
\begin{equation}\label{I1main}
  I_1(\bx)=\frac{\pi^{\frac52}}{K(\frac1{\sqrt2})}\,r^{-\frac32}
  {}_2F_1\biggl(\frac34,\frac12;1;\sin^2\th\biggr).
\end{equation}
Finally, applying the identity~\eqref{2F1tr1} and eq.~\eqref{EK2F1} we arrive at the following
more compact expression:
\begin{equation}\label{Ibe1}
  I_1(\bx)=
  (2\pi)^{\frac32}r^{-\frac32}s^{-\frac12}(1+s)^{-\frac12}
  \frac{K(\ka_1)}{K(\frac1{\sqrt2})},
\end{equation}  
where $s=|\cos\th|$ and $\ka_1$ is given by eq.~\eqref{ka1}.
Note that $0\le \ka_0,\ka_1\le 1$, so that the elliptic integrals are real for all $\bx\in\RR$.
Substituting eqs.~\eqref{Ibe0} and \eqref{Ibe1} into eq.~\eqref{1.3}, and using the standard
formulas for the derivatives of the complete elliptic integrals $K(\ka)$ and $E(\ka)$ listed in
eqs.~\eqref{Kp}-\eqref{Ep}, after a straightforward calculation we obtain
\begin{equation}
  \label{bPsi01}
  \bPsi_{\bm 0}^{(1)}(\bx)=\sqrt2\,(\hbar c)^{\frac\be2} r^{-3-\frac\be2}\big(P_\rho(\th)\bm e_\rho+P_z(\th)\bm e_z\big),
\end{equation}
with $P_\rho$ and $P_z$ given by eqs.~\eqref{LPPs}-\eqref{RSPs}.

\subsection{The function \texorpdfstring{$\bPsi_{\bm0}^{(2)}(\bx)$}{\relax}}
\smallskip\noindent
From the expression of $\bm{E}_2$ in eq.~\eqref{bE2} and eq.~\eqref{crj} we have
\begin{align}
  i\,\bPsi_{\bm{0}}^{(2)}(\bx)
  &=-(2\pi)^{-\frac32}i\mkern-6mu\int_{\mathbb{R}^3} \dd^3k\,( \bk\times
    \bm{e}_3) \frac{k^{\frac\be2}}{k_{\perp}}e^{i \bk \cdot \bx }\notag\\
  &=\frac{(\hbar c)^{\be/2}}{(2\pi)^{\frac32}}\,\bm{e}_3\times\bm{\nabla}I_{\be+2}(\bx).
\label{gradJ}
\end{align}
By eq.~\eqref{main} with $\be$ replaced by $\be+2$ we then have
\begin{align}
  I_{\be+2}(\bx)
  &=\pi^2\frac{\Ga(1+\frac\be4)}{\Ga(-\frac\be4)}\,\left(\frac2r\right)^{\frac\be2+2}
    {}_2F_1\left(1+\frac\be4,\frac12;1;\sin^2\th\right)\notag\\
  &=-2^{\frac\be2}\be\pi^2\frac{\Ga(1+\frac\be4)}{\Ga(1-\frac\be4)}\,r^{-\frac\be2-2}
    {}_2F_1\left(1+\frac\be4,\frac12;1;\sin^2\th\right).
    \label{Ibep2}
\end{align}
As explained above, for $k_\perp\ne0$ the integral defining $I_{\be+2}$ converges only for
$\be<0$. In particular, it is divergent for both $\be=0$ and $\be=1$. However, analytically
continuing it to $\be=1$ and using the identity
\[
  \frac{\Ga(\frac54)}{\Ga(\frac34)}=\frac14\frac{\Ga(\frac14)}{\Ga(\frac34)}=
  \frac{\Ga(\frac14)^2}{4\sqrt2\,\pi}=\frac{K(\frac1{\sqrt2})}{\sqrt{2\pi}}
\]
we obtain the regularized value
\begin{equation}\label{Jbe1}
  I_3(\bx)=-\pi^{\frac32}K\bigl(\tfrac1{\sqrt2}\bigr)r^{-\frac52}
  {}_2F_1\left(\frac54,\frac12;1;\sin^2\th\right).
\end{equation}
Transforming the last equation with the help of the identity~\eqref{2F1tr2} we finally have
\begin{multline*}
  I_3(\bx)=-(8\pi)^{\frac12}K(\tfrac1{\sqrt2})\,|x_3|^{-\frac52}\frac{s}{\sqrt{1+s}}\,
  \big[2(1+s)E(\ka_1)\\
    -(s+\sqrt{s}+1)K(\ka_1)\big],
\end{multline*}
where again $s$ and $\ka_1$ are as above. Substituting into eq.~\eqref{gradJ} with $\be=1$, and
taking into account the identity
\begin{align*}
  \pdf{}{x_j}f(s,|x_3|)&=\pdf{f}{s}(s,|x_3|)\pdf{s}{x_j}\\
    &=-\frac{|x_3|}{r^3}\,x_j\pdf{f}{s}(s,|x_3|),\quad
  j=1,2,
\end{align*}
after a long but straightforward calculation we obtain
\begin{equation}
  \label{bPsi02}
  i\cF_{\bm 0}^{(2)}(\bx)=\sqrt2\,(\hbar c)^{\frac12}r^{-\frac72}P_\psi(\th)\bm e_\psi,
\end{equation}
with $P_\psi$ given by eq.~\eqref{RSPs}.
  
In order to derive the explicit expression for the LP eigenfunction $\bPsi_{\bm0}^{(2)}$, it is
better to start directly from eq.~\eqref{1.2} with $\be=2$ and evaluate the integral with respect
to $k_3$ in terms of a delta function. We thus obtain:
\begin{align*}
  I_2(\bx)&=\int_{\mathbb{R}^2} \dd^2k_{\perp}\,
    \frac{ e^{i \bk_{\perp} \cdot \bx_{\perp} }}{k_{\perp}}
    \int_{-\infty}^{\infty} \dd k_3\, e^{i k_3  x_3}\\
      &=2\pi\de(x_3)\int_{\mathbb{R}^2} \dd^2k_{\perp}\,
    \frac{ e^{i \bk_{\perp} \cdot \bx_{\perp} }}{k_{\perp}}.
\end{align*}
The last integral can be computed in polar coordinates \mbox{$(\rho=k_\perp,\phi)$} for $\bk$ and
$(x_{\perp}, \psi)$ for $\bx_{\perp}$ with the help of eqs.~\eqref{A3}, with the result
\begin{align*}
  I_2(\bx)
  &=
    2\pi\de(x_3)\int_0^\infty\dd\rho\,\int_0^{2\pi}\dd\phi\,e^{i\rho x_\perp\cos(\phi-\psi)}\\
  &=4\pi^2\de(x_3)\int_0^\infty\dd\rho\,J_0(\rho x_\perp)
  =\frac{4\pi^2}{\rho}\,\de(x_3).
\end{align*}
Substituting into eq.~\eqref{gradJ} with $\be=0$ we finally obtain
\begin{equation}
  \label{bPsi02LP}
  i\bPsi_{\bm0}^{(2)}(\bx)=\sqrt2\,r^{-3}P_\psi(\th)\bm e_\psi,
\end{equation}
with $P_\psi$ given by eq.~\eqref{LPPs}.

\section{calculation of the Hertz superpotential}\label{app.Hertz}

  By eqs.~\eqref{bPsisi}, the RS position eigenfunction with helicity $\si$ is given by
  \begin{multline}\label{he 7}
    \cF_\si(\bx,t)\\
    =\int_{\RR^3}\frac{\dd^3k}{(2\pi)^{\frac32}}\,\bigg(\frac{\hbar ck}{2}\bigg)^{\frac12}
    \Big(\bm{E}_1(\bk) +i\si \bm{E}_2(\bk) \Big)   
    e^ {i(\bm{k\cdot \bm{x}-\omega t)}}.
  \end{multline}
  From the explicit expressions~\eqref{basis} for the vectors $\bm{E}_i$ $(i=1,2)$ of the standard
  orthonormal frame in spherical coordinates and eqs.~\eqref{he3}, it then follows that the
  Fourier transform $\bh(\bk)$ of the Hertz superpotential $\bZ$ for $\cF_\si$ at $t=0$ is given
  by
\begin{equation}\label{he8}
  \bh(\bm{k})=- \bigg(\frac{\hbar c}{2}\bigg)^{\frac12}\frac{\si}{k^{\frac12} k_{\perp}}\,\bm{e}_3.
\end{equation}
We thus have
\begin{align}\label{he9}
  \bm {Z}(\bx,0)
  &=-\si\bigg(\frac{\hbar c}{2}\bigg)^{\frac12}\bm e_3\int_{\RR^3}\frac{\dd^3k}{(2\pi)^{\frac32}}
    \frac{e^{i\bk\cdot\bx}}{k_\perp k^{\frac12}}\\
  &=-\frac{\si}{(2\pi)^{\frac32}}\bigg(\frac{\hbar c}{2}\bigg)^{\frac12}I_1(\bx)\,\bm e_3,
\end{align}
with $I_1$ defined by eq.~\eqref{1.2} with $\be=1$. From eqs.~\eqref{I1main} we then obtain
eq.~\eqref{Zx0} for $\bZ(\bx,0)$.

We have not been able to derive a closed-form expression for the Hertz superpotential at an
arbitrary time $t$,
\begin{equation}\label{Hertzt}
  \bZ(\bx,t)=-\si\bigg(\frac{\hbar c}{2}\bigg)^{\frac12}
  \bm e_3\int_{\RR^3}\frac{\dd^3k}{(2\pi)^{\frac32}}
  \frac{e^{i(\bk\cdot\bx-ckt)}}{k_\perp k^{\frac12}}.
\end{equation}
It is, however, straightforward to obtain the Taylor series of this function about $t=0$. Indeed,
differentiating the last expression with respect to $t$ we obtain:
\begin{align*}
  (\pd_t^n\bZ)(\bx,0)
  &=-(-i c)^n\si\bigg(\frac{\hbar c}{2}\bigg)^{\frac12}
    \bm e_3\int_{\RR^3}\frac{\dd^3k}{(2\pi)^{\frac32}}
    \frac{k^{n-\frac12}}{k_\perp}e^{i\bk\cdot\bx}\\
  &=-\si(2\pi)^{-\frac32}(-i c)^n\bigg(\frac{\hbar c}{2}\bigg)^{\frac12} I_{2n+1}\,
    \bm e_3,
\end{align*}
where $I_{2n+1}$ is the integral in eq.~\eqref{1.2} with $\be=2n+1$. Using the explicit
formula~\eqref{main} for this integral we obtain
\[
  \bZ(\bx,t)=-\si(\pi\hbar c)^{\frac12}\ze(\bx,t)\,\bm e_3,
\]
with
\begin{widetext}
  \[
    \ze(\bx,t)=\frac14\sum_{n=0}^\infty \frac{\Ga(\frac n2+\frac34)}{\Ga(\frac14-\frac
      n2)}\,\left(\frac2r\right)^{n+\frac32} {}_2F_1\left(\frac
      n2+\frac34,\frac12;1;\sin^2\th\right) \frac{(-ict)^n}{n!}.
  \]
\end{widetext}
Applying the reflection property~\eqref{gammar} of the gamma function and the fundamental identity
$\Ga(z+1)=z\Ga(z)$ we can simplify the quotients of gamma functions in the previous series as
follows:
\begin{widetext}
  \begin{equation*}
    \frac{\Ga(\frac n2+\frac34)}{\Ga(\frac14-\frac n2)}
    =(-1)^{\frac12(n+\pi(n))}\left(\frac{K(\frac1{\sqrt2})}{\sqrt{2\pi}}\right)4^{-(n+1-\pi(n))}\\
    \prod_{j=0}^{\frac12(n-\pi(n))}\big(4j-1+2\pi(n)\big)^2,
  \end{equation*}
\end{widetext}
where $\pi(n)=n\pmod 2\in\{0,1\}$ is the parity of $n$. Substituting this expression into the
previous expansion we obtain the more explicit formula~\eqref{Zseries}. Note, finally, that the
hypergeometric functions appearing in the Taylor expansion of $\ze$ can be expressed in terms of
the complete elliptic integrals $K(\ka_1)$ and $E(\ka_1)$, with $\ka_1$ given by eq.~\eqref{ka1}
and $s=|\cos\th|$. Indeed, taking into account that $x_\perp^2/r^2=1-s^2$, and applying
eqs.~\eqref{2F1tr1}-\eqref{2F1tr2} and eq.~\eqref{2F1ap1}, we readily obtain
\begin{widetext}
  \begin{align*} {}_2F_1\left(k+\frac34,\frac12;1;\frac{x_\perp^2}{r^2}\right)
    &=\frac{2^{\frac32}}{\pi}\left(1-\frac23\frac{1-s^2}{s}\pd_s\right)^k
      \frac{K(\ka_1)}{\sqrt{s(1+s)}},\\
    {}_2F_1\left(k+\frac54,\frac12;1;\frac{x_\perp^2}{r^2}\right)
    &=\frac{2^{\frac32}}{\pi}\left(1-\frac25\frac{1-s^2}{s}\pd_s\right)^k
      \left[s^{-3/2}(1+s)^{-1/2}\Big( 2(1+s)E(\ka_1)-(s+\sqrt s+1)K(\ka_1))\Big) \right].
  \end{align*}
\end{widetext}


%

\end{document}